\documentclass[12pt]{article}

\usepackage{graphicx,epsfig,rotate,amsmath}    
\usepackage{latexsym,color,amssymb,citesort}
\newcommand{\be}{\begin{equation}}
\newcommand{\ee}{\end{equation}}
\newcommand{\eq}{\begin{eqnarray}}
\newcommand{\en}{\end{eqnarray}}
\newcommand{\bea}{\begin{eqnarray}}
\newcommand{\eea}{\end{eqnarray}}

\newcommand{\ed}{\end{document}}
\newcommand{\bc}{\begin{center}}
\newcommand{\ec}{\end{center}}

\begin{document}

\thispagestyle{empty}

\begin{flushright}
{\tiny Preprint HISKP--TH--08/08, FZJ-IKP(TH)--08--10}
\end{flushright}

\begin{center}

\vspace{1.75cm}
{\Large{\bf Resonance properties from the finite-volume\\[0.3em]
 energy spectrum}\footnote{This
research is part of the EU Integrated Infrastructure Initiative Hadron Physics Project
under contract number RII3-CT-2004-506078. Work supported in part by DFG (SFB/TR 16,
``Subnuclear Structure of Matter''), by the Helmholtz Association
through funds provided to the virtual institute ``Spin and strong
QCD'' (VH-VI-231) and by COSY FEE grant ``Inelastic baryon resonances from lattice QCD'' under contract number 41821485 (COSY-106).}}

\vspace{0.5cm}

\today

\vspace{0.5cm}

V. Bernard$^a$,
M. Lage$^b$,
U.-G. Mei{\ss}ner$^{b,c}$ and
A.~Rusetsky$^b$

\vspace{2em}

\begin{tabular}{c}
$^a\,${\it Universit\'e Louis Pasteur, Laboratoire de Physique
            Th\'eorique}\\ {\it 3-5, rue de l'Universit\'e,
            F--67084 Strasbourg, France}\\[2mm]
$^b\,${\it Helmholtz--Institut f\"ur Strahlen-- und Kernphysik}\\
{\it Universit\"at Bonn, Nu{\ss}allee~14--16, D--53115 Bonn, Germany}\\[2mm]
$^c${\it Forschungszentrum J\"ulich, Institut f\"ur Kernphysik 
(Theorie)}\\ {\it D-52425 J\"ulich, Germany}
\end{tabular}

\end{center}

\vspace{1cm}

{\abstract
{A new method based on the concept of probability distribution is 
proposed to analyze the finite volume energy spectrum in lattice QCD. 
Using synthetic lattice data, we demonstrate that for the channel 
with quantum numbers of the $\Delta$-resonance
a clear resonance structure emerges in such an analysis.
Consequently, measuring the volume-dependence of the energy levels in 
lattice QCD will allow to determine the mass and the width of the $\Delta$ 
with reasonable accuracy.
}
}

\vskip1cm

{\footnotesize{\begin{tabular}{ll}
{\bf{Pacs:}}$\!\!\!\!$& 11.10.St, 11.15.Ha\\
{\bf{Keywords:}}$\!\!\!\!$& Resonances in lattice QCD,
field theory in a finite volume,\\$\!\!\!\!$ &non-relativistic EFT
\end{tabular}}
}
\clearpage


\section{Introduction}

Recent years have seen a substantial growth of interest in the 
calculation of excited baryon spectra in lattice 
QCD~\cite{Richards:2001bx,Maynard:2002ys,Gattringer:2003qx,Sasaki:2001nf,Zanotti:2001yb,Lasscock:2007ce,Melnitchouk:2002eg,Zhou:2006xe,Mathur:2003zf,Guadagnoli:2004wm,Alexandrou:2007qq} that
has largely been motivated by the present experimental programs at 
Jefferson Lab~\cite{Jlab} and ELSA~\cite{BeckThoma} 
(for the latest lattice reviews, see
e.g.~\cite{McNeile:2003dy,Leinweber:2004it,Michael:2005kw,Gattringer:2007da}).
In general, the procedure of extracting  resonances from lattice
QCD data differs from the one used in  the stable particle case, since 
resonances do not correspond to isolated energy levels of
the total Hamiltonian. The standard approach, originally proposed by 
L\"uscher~\cite{LuescherII,Luescher_torus,Luescher_rho,Houches}
(see also \cite{Baal,Wiese,DeGrand}),
is based on studying the volume dependence of the spectrum being
determined by the two-body scattering phase shift in the infinite volume. 
Near the resonance energy, where the phase shift rapidly passes through
$\pi/2$, an abrupt rearrangement of the energy levels known as
``avoided level crossing'' takes place. It has been argued that
the observation of this phenomenon in lattice data can serve as
a signal of the presence of a resonance and enables one to determine
its parameters. In Refs.~\cite{Rummukainen:1995vs,Christ:2005gi,Kim:2005gf} 
the approach has been further generalized for moving frames.
Note also that  L\"uscher's approach has been recently applied 
to study nucleon-nucleon phase shifts at low-energy, as well as two-body 
shallow bound 
states~\cite{Beane:2003yx,Beane:2003da,Beane:2006gf,Beane:2006mx,Sasaki:2006jn}.

As alternative approaches to this procedure we mention, e.g.,
Ref.~\cite{Michael}, where it has been shown that the presence of a 
narrow excited state above the threshold modifies the simple exponential decay
law of the time-sliced two-point function. The decay width
within this approach is extracted not from the two-point function, but 
directly from the decay amplitudes 
(see also~\cite{Loft:1988sy,Lellouch:2000pv}).
 In addition, in Ref.~\cite{Yamazaki} it was proposed to reconstruct the 
spectral density in the two-point function by using the maximum entropy method.
This approach, in principle, also has the capability to address the
problem of unstable states in lattice calculations.

It turns out that the lowest-mass strongly interacting unstable particles 
in nature, the $\rho (770)$, 
the $\Delta (1232)$-resonance, etc~\footnote{We eschew here the $\sigma (600)$
since at present there is no consensus about the precise nature of this
resonance.}, due to their large width can not be identified 
with a clearly visible
bifurcation in the energy levels. Namely, predicting
the volume dependence of the pertinent energy levels
 by using the experimentally measured $\pi\pi$ and 
$\pi N$ phase shifts, it is seen that the avoided level crossing is almost
completely washed out. In this case, it is natural to ask, whether
the resonance parameters which will be extracted by fitting  L\"uscher's 
formula to the lattice data, can be determined at a reasonable accuracy 
and will be devoid of any bias. The issue of accuracy becomes 
particularly important since in forthcoming lattice calculations 
resonance parameters will be fitted to a few available data points.
For example, in Ref.~\cite{Aoki:2007rd} 
the parameters of the $\rho$-meson have been determined 
performing a fit of L\"uscher's formula (in a moving frame) 
only to two data points. We believe that with more data on resonances expected to come, 
this question should be urgently addressed.

In a previous paper~\cite{delta1} we have studied the problem in the
case of the $\Delta$-resonance. Invoking chiral effective field theory 
with explicit spin-3/2 degrees of freedom,
we have parameterized the volume-dependent energy spectrum
of the total Hamiltonian in terms of the $\Delta$-resonance mass and width
up to third order in the so-called small scale
expansion (see, e.g.~\cite{Jenkins:1991es,Hemmert:1997ye}).
On the basis of a detailed analysis of the behavior of the two 
lowest energy levels, it was
concluded that an accurate extraction of the $\Delta$-resonance parameters is
indeed a feasible task, despite the fact that the avoided level crossing
is completely washed out.

In the present paper we address the same problem within non-relativistic
effective field theory (NR EFT) in a finite volume, which enables 
one to carry out
the analysis in a more general way\footnote{Some of the results of this paper
have been reported previously~\cite{NSTAR}.}. 
The equation that determines the location of the eigenvalues 
of the Hamiltonian in this framework coincides with L\"uscher's formula.
In order to facilitate the analysis,
we further define a so-called probability distribution, which is constructed
from the volume-dependent energies.
The central observation is that the
probability distribution in the vicinity of a resonance behaves much like the
infinite-volume scattering cross section: it peaks at the resonance energy.
The peak has approximately a Breit-Wigner shape, with the same width as
the original resonance. We will show in the following that in case of a wide resonance, 
when the avoided level crossing is washed out, one still observes a clear 
resonance structure in the probability distribution after subtracting the 
background corresponding to the free motion of the decay products. This 
result unanimously supports the conclusion of Ref.~\cite{delta1}: the 
extraction of both the energy and width of the $\Delta$-resonance
from the volume-dependent spectrum by using L\"uscher's formula
is feasible. Note also that, as shown in 
the present paper, this goal can be achieved even by fitting to 
the data for the lowest energy level alone. 

An important issue, which we do not discuss in the present paper,
concerns the quark mass dependence of the resonance observables
and the energy spectrum. For the values of the pion masses, which are 
used in present day  calculations, the volume dependence of the energy levels 
may qualitatively differ from what happens at the physical value of the 
pion mass. For example, at higher pion masses the $\Delta$-resonance becomes
stable and the peak in the probability distribution degenerates into a 
$\delta$-function. Our approach, combined with chiral perturbation theory,
is flexible enough to describe this continuous transition. We, however, relegate
a detailed discussion of this question to a separate publication~\cite{Hoja}.

The layout of the present paper is the following. In section~\ref{sec:NR} we 
consider the NR EFT for the $\pi N$ system 
and derive L\"uscher's formula for spin-$0$ particle scattering on a spin-1/2 particle. 
Section~\ref{sec:cubic} contains the reduction of  
L\"uscher's formula, using the cubic symmetry of the lattice. In 
section~\ref{sec:probability} we introduce the notion of the probability
distribution and discuss its properties, as well as the infinite-volume limit. 
Finally, section~\ref{sec:analysis} contains an analysis of 
(synthetic) lattice data, performed with the use of the probability 
distribution technique. We end with a summary and conclusions in 
sec.~\ref{sec:concl}. Some technicalities are relegated to the appendices.

\section{Non-relativistic EFT for the pion-nucleon system
in a finite volume}
\label{sec:NR}

L\"uscher's formula, which relates the infinite-volume elastic phase shift
to the finite-volume two-particle energy spectrum, is derived 
in large volumes. Namely, the size of the three-dimensional box $L$, in which
the two-particle system is placed, 
should be much larger than the typical scale 
$M_{\pi}^{-1}$ set by the mass of the lightest particle (the pion in our case),
in order to be able to discard all exponentially suppressed contributions
at large $L$.
For such large volumes, NR EFT provides an adequate description of the
system at low energies. L\"uscher's formula within NR EFT can be straightforwardly obtained 
(see, e.g.~\cite{Beane:2003yx}). In this
section we briefly describe the generalization of the method to the case of 
particles with spin. Although the approach is completely general,
below we shall focus on the example of pion-nucleon scattering.
We shall use the covariant formulation of the NR EFT, introduced in 
Refs.~\cite{cuspwe}. The relativistic kinematics is taken into account 
automatically in this formulation that, in particular, may prove advantageous
for generalizing L\"uscher's approach to moving frames. A recent general
introduction to the NR EFT can be found, e.g. in Ref.~\cite{Gasser:2007zt}.

The derivation consists of two parts. In the first part we set up the framework
by considering the pion-nucleon scattering process in dimensionally 
regularized NR EFT in the infinite volume. To ease the notation, we suppress
the isospin indices everywhere in the following, considering 
the scattering process in a channel with fixed total isospin. The 
non-relativistic Lagrangian takes the form
\eq\label{eq:L}
{\cal L}=\Phi^\dagger\,2W_\pi(i\partial_t-W_\pi)\Phi
+\Psi^\dagger\,2W_N(i\partial_t-W_N)\Psi+{\cal L}_I\, ,
\en
where $\Phi$ and $\Psi$ denote the non-relativistic pion and nucleon fields,
respectively. Further,  $W_\pi=(M_\pi^2-\nabla^2)^{1/2}$ and  
$W_N=(m_N^2-\nabla^2)^{1/2}$, with $M_\pi$ and $m_N$ the physical masses of the
pion and the nucleon, in order. The $\pi N$ interaction Lagrangian 
${\cal L}_I$ contains a tower of local 4-particle operators with increasing 
powers of space derivatives. The number of heavy particles is conserved.
The coupling constants in ${\cal L}_I$
encode the whole information about the high-energy behavior of the theory and 
are determined through matching to the effective-range expansion of the
physical amplitudes.

The non-relativistic pion and nucleon propagators are given by
\eq
S_\pi(p)=\frac{1}{2w_\pi({\bf p})}\,\frac{1}{w_\pi({\bf p})-p^0-i0}\, 
,\quad\quad
S_N(p)=\frac{1}{2w_N({\bf p})}\,\frac{1}{w_N({\bf p})-p^0-i0}\, ,
\en
where $w_\pi({\bf p})=(M_\pi^2+{\bf p}^2)^{1/2}$ and
$w_N({\bf p})=(m_N^2+{\bf p}^2)^{1/2}$.

The Lagrangian in Eq.~(\ref{eq:L}) generates loops through the 
usual Feynman diagrammatic technique. 
In order to ensure power counting and relativistic covariance,
Feynman rules are supplemented with an additional prescription~\cite{cuspwe}:
the integrands in all Feynman integrals are expanded in the inverse powers
of masses, integrated by using dimensional regularization and finally summed
up again to all orders. In the two-particle sector, which is considered here,
 this procedure is 
straightforward, since all loop contributions can be expressed through
the basic bubble integral
\eq\label{eq:bubble}
J(s)&=&
-i \int\frac{d^Dl}{(2\pi)^D}\,\frac{1}{2w_\pi({\bf l})2w_N({\bf P}-{\bf l})}
\,\frac{1}{(w_\pi({\bf l})-l^0)(w_N({\bf P}-{\bf l})-P^0+l^0)}
\nonumber\\[2mm]
&=&\frac{iq(s)}{8\pi\sqrt{s}}+O(d-3)\, ,\quad\quad
q(s)=\frac{\lambda^{1/2}(s,M_\pi^2,m_N^2)}{2\sqrt{s}}\, ,\quad s=P^2\, ,
\en
where $\lambda(x,y,z)=x^2+y^2+z^2-2xy-2yz-2zx$ denotes the triangle function,
$D$ is the number of space-time dimensions and $d=D-1$.

Using canonical formalism, the full Hamiltonian of the $\pi N$ system
${\bf H}={\bf H}_0+{\bf H}_I$ can be constructed from the 
Lagrangian~(\ref{eq:L}). The scattering matrix ${\bf T}(z)$ is defined through
the Lippmann-Schwinger (LS) equation
\eq\label{eq:LS}
 {\bf T}(z)=(-{\bf H}_I)+(-{\bf H}_I)(-{\bf G}_0(z)){\bf T}(z)\, ,
\en
where ${\bf G}_0(z)=(z-{\bf H}_0)^{-1}$ is the free resolvent of the 
$\pi N$ system. Note that we have chosen to introduce
 negative signs in the above equation, in order
to take into account the different sign conventions
for the $T$-matrix in the field theory and in the potential scattering theory.

Next, we define the center-of-mass (CM) and relative momenta of the 
pion-nucleon pair, {\bf P} and {\bf k}, respectively,
\eq
{\bf p}_N=\frac{m_N}{m_N+M_\pi}\,{\bf P}+{\bf k}\, ,\quad
{\bf p}_\pi=\frac{M_\pi}{m_N+M_\pi}\,{\bf P}-{\bf k}\, .
\en
The pion-nucleon states are given by
\eq
|({\bf p}_N\nu),{\bf p}_\pi\rangle=|{\bf P},{\bf k},\nu\rangle\, ,
\en
where the index $\nu$ labels the nucleon spin. The normalization of
the states is fixed by
\eq
\langle {\bf P}',{\bf k}',\nu'|{\bf P},{\bf k},\nu\rangle
=2w_N({\bf p}_N)2w_\pi({\bf p}_\pi)\delta_{\nu'\nu}
(2\pi)^d\delta^d({\bf P}'-{\bf P})(2\pi)^d\delta^d({\bf k}'-{\bf k})\, .
\en
We remove the CM momentum in the matrix elements by defining
\eq
 t_{\nu' \nu}(\mathbf{k}', \mathbf{k}; z )&=& 
\int\frac{d^d{\bf P}'}{(2\pi)^d}\,
\langle {\bf P}',{\bf k}',\nu'| {\bf T}(z) |{\bf 0},{\bf k},\nu\rangle\, , 
\nonumber\\[2mm]
 h_{\nu' \nu}(\mathbf{k}', \mathbf{k})&=& 
-\int\frac{d^d{\bf P}'}{(2\pi)^d}\,
\langle {\bf P}',{\bf k}',\nu'| {\bf H} |{\bf 0},{\bf k},\nu\rangle \, .
\en
The LS equation in the CM frame takes the form
\begin{eqnarray}
 t_{\nu' \nu}(\mathbf{k}', \mathbf{k}; z )=h_{\nu' \nu}(\mathbf{k}', \mathbf{k})
+  \sum_{\nu^{\prime\prime}}  \int \frac{d^d k^{\prime\prime}}{(2\pi)^d} 
                               \frac{1}{2 \omega_N(\mathbf{k}^{\prime\prime})} 
                               \frac{1}{2 \omega_\pi(\mathbf{k}^{\prime\prime})}  \,\,\frac{h_{\nu' \nu^{\prime\prime}}(\mathbf{k}', \mathbf{k}^{\prime\prime})
      t_{\nu^{\prime\prime} \nu}(\mathbf{k}^{\prime\prime}, \mathbf{k}; z )}{\omega_N(\mathbf{k}^{\prime\prime})+\omega_\pi(\mathbf{k}^{\prime\prime})-z}  \label{LS-eqn}
      ~.
\end{eqnarray}
Since $h_{\nu'\nu}({\bf k}',{\bf k})$ is (an infinite)
 polynomial in momenta, by using
Eq.~(\ref{eq:bubble}) the above equation simplifies to
\eq\label{eq:onshell}
 t_{\nu' \nu}(\mathbf{k}', \mathbf{k}; z )
=h_{\nu' \nu}(\mathbf{k}', \mathbf{k})
+\frac{iq(z^2)}{32\pi^2z}\, \sum_{\nu^{\prime\prime}}
\int d\Omega_{{\bf k}^{\prime\prime}}\,
h_{\nu' \nu^{\prime\prime}}({\bf k}',\tilde{\bf k}^{\prime\prime})\,
t_{\nu^{\prime\prime} \nu}(\tilde {\bf k}^{\prime\prime}, {\bf k}; z )\, ,
\en
where $d\Omega_{{\bf k}^{\prime\prime}}$ stands for the integral over the 
3-dimensional solid angle and
\eq
\tilde{\bf k}^{\prime\prime}=
\frac{{\bf k}^{\prime\prime}}
{|{\bf k}^{\prime\prime}|}\,
\frac{\lambda^{1/2}(z^2,M_\pi^2,m_N^2)}{2z}\, .
\en
Note that, since the pion-nucleon loop in Eq.~(\ref{eq:bubble}) is 
finite at $d\to 3$, one may set $d=3$ in Eq.~(\ref{eq:onshell}).
All terms that vanish in dimensional regularization after performing the
expansion in the inverse powers of masses are disregarded. 
The partial-wave
expansion proceeds then in the standard manner. It is carried out in terms
of spinor spherical harmonics defined by
\begin{equation}\label{Y_JLM}
 {\cal Y}_{J M}^{L \frac{1}{2}}(\hat{\mathbf{k}},\nu)=\sum_{m_l m_s} 
\langle L m_l\,\frac{1}{2}\, m_s \vert J M \rangle
              Y_{m_l}^{L}(\hat{\mathbf{k}}) \chi_{m_s}(\nu)~,
\end{equation}
where $\hat{\mathbf{k}}=\mathbf{k}/|{\bf k}|$ and 
$Y_{lm}(\hat{\bf k})$ and $\chi_{m_s}(\nu)$ are the Legendre spherical function
and the two-component nucleon spinor, respectively. The quantity
$\langle L m_l\,\frac{1}{2}\, m_s \vert J M \rangle$ stands for the pertinent 
Clebsch-Gordan coefficient.

\noindent
Introducing for convenience the projectors
\eq\label{PW-operators}
 \Pi^{J L}_{\nu' \nu}(\hat{\mathbf{k}}',\hat{\mathbf{k}})
=\sum_M  
({\cal Y}_{J M}^{L \frac{1}{2}}(\hat{\mathbf{k}'},\nu'))^*
 {\cal Y}_{J M}^{L \frac{1}{2}}(\hat{\mathbf{k}},\nu)\, ,
\en
the partial wave expansion can be written as
\begin{eqnarray}
 t_{\nu' \nu}(\mathbf{k}', \mathbf{k}; z)&=& 4\pi \sum_{J L}
                    \Pi^{J L}_{\nu' \nu}(\hat{\mathbf{k}}',\hat{\mathbf{k}}) 
                              t_{J L}(k',k; z)~, \nonumber\\[2mm]
 h_{\nu' \nu}(\mathbf{k}', \mathbf{k}) &=& 4 \pi \sum_{J L}
                    \Pi^{J L}_{\nu' \nu}(\hat{\mathbf{k}}',\hat{\mathbf{k}}) 
                              h_{J L}(k',k)~. \label{PW-expansion}
\end{eqnarray}
On the mass shell, $k=k'$ and $z(k)=(M_\pi^2+k^2)^{1/2}+(m_N^2+k^2)^{1/2}$,
the scattering amplitude and the matrix element of the Hamiltonian
can be expressed through the elastic scattering phases $\delta_{JL}(k)$
in a standard manner
\begin{eqnarray}
 t_{J L}(k,k;z(k))&=&\frac{8 \pi z(k)}{k}~ \frac{\exp\left( 2i \delta_{J L}(k) \right)-1}{2i}~,\nonumber\\[2mm]
 h_{J L}(k,k)&=&\frac{8 \pi z(k)}{k}~ \tan \delta_{J L}(k)~. \label{H-by-phase}
\end{eqnarray}
At the next step we consider the same system placed in a finite cubic box
$L\times L\times L$. The Feynman rules in a finite volume remain the same,
except that the momentum integration everywhere is now replaced by a 
discrete sum
\begin{equation}
 \int \frac{d^d k^{\prime\prime}}{(2 \pi)^d} \rightarrow \frac{1}{L^3} 
     \sum_{\mathbf{k}^{\prime\prime}},~~~~\mathbf{k}^{\prime\prime}=\frac{2\pi \mathbf{n}}{L},~\mathbf{n} \in \mathbb{Z}^3~. \label{sums}
\end{equation}
Our aim is to find the finite-volume energy spectrum
of the system described by the Lag\-ran\-gi\-an~(\ref{eq:L}). 
To this end, note that the location 
of the eigenvalues of the Hamiltonian in a finite volume coincides with 
the (real) poles of the operator ${\bf T}(z)$, defined by Eq.~(\ref{eq:LS}), in
a complex $z$-plane. After removing the CM momentum, this equation becomes
similar to Eq.~(\ref{LS-eqn}), but with the integration replaced through the
momentum sum, as in Eq.~(\ref{sums}). The ultraviolet divergence can be 
most conveniently tamed by analytic regularization~\cite{Luescher_torus}.
In the following, we do not indicate the regularization explicitly.

The resulting equation can again be expanded in partial waves. 
Since the rotational symmetry in the infinite volume is 
now broken down to a cubic symmetry, the partial wave expansion of the matrix
elements of the operator ${\bf T}(z)$ will 
not be diagonal in $J$,$L$ and $M$ anymore. In order to ease the notations,
it is useful to introduce the multi-index $A=(J,L,M)$.
Defining the operators
\begin{eqnarray}
 \Pi^{A'A}_{\nu' \nu}(\hat{\mathbf{k}}',\hat{\mathbf{k}})
   &\doteq&({\cal Y}_{J' M'}^{L' \frac{1}{2}}(\hat{\mathbf{k}'},\nu'))^*
 {\cal Y}_{J M}^{L \frac{1}{2}}(\hat{\mathbf{k}},\nu)\, ,
                                     \label{PW-operators-finite}
\end{eqnarray}
the partial wave expansion can be written as
\begin{eqnarray}
 t_{\nu' \nu}(\mathbf{k}', \mathbf{k}; z)&=& 4\pi 
                               \sum_{ A' A }
               \Pi^{A' A}_{\nu' \nu}(\hat{\mathbf{k}}',\hat{\mathbf{k}})
                              t_{A' A}(k',k; z)~, \nonumber\\[2mm]
 h_{\nu' \nu}(\mathbf{k}', \mathbf{k}) &=& 4 \pi 
                               \sum_{ A' A }
                    \Pi^{A' A}_{\nu' \nu}(\hat{\mathbf{k}}',\hat{\mathbf{k}})
h_{A' A}(k',k)
~.
 \label{PW-expansion-finite}
\end{eqnarray}
Further, since $h_{A' A}(k',k)$ is calculated from the non-relativistic 
Lagrangian at tree level, it coincides with its infinite-volume counterpart
and is diagonal
\eq
h_{A' A}(k',k)=\delta_{A'A}h_A(k',k)
=h_{J L}(k',k) \delta_{J' J} \delta_{L' L} \delta_{M' M}\, .
\en
Next, in analogy with the infinite-volume case,
one may  derive the equation for the on-shell quantities. To this end, note
that the off-shell contribution to the LS equation is exponentially suppressed
by the box size $L$, since the singular energy denominator
is canceled in this contribution 
(for the proof of this statement, see, e.g.~\cite{Christ:2005gi}). Thus, up
to these exponentially suppressed terms, the partial-wave expanded LS equation in a finite volume can
be rewritten as
\begin{eqnarray}\label{eq:LSfin}
&&\hspace*{-.43cm}t_{A'A}(k,k;z(k))- \delta_{A'A}h_A(k,k)=\frac{k}{8\pi z(k)}\sum_{A''}
h_{A'}(k,k){\cal M}_{A'A''}(k)t_{A''A}(k,k;z(k))\, ,
\nonumber\\
&&
\end{eqnarray}
where we have defined
\eq\label{eq:M}
{\cal M}_{A'A}(k)={\cal M}_{J'L'M',JLM}(k)=\frac{16\pi^2}{k}\,
\frac{1}{L^3}\sum_{{\bf k}^{\prime\prime}}\sum_\nu
\frac{({\cal Y}_{J' M'}^{L' \frac{1}{2}}(\hat{\mathbf{k}^{\prime\prime}},\nu))^*
 {\cal Y}_{J M}^{L \frac{1}{2}}(\hat{\mathbf{k}^{\prime\prime}},\nu)}{{\bf k^{\prime\prime}}^2-k^2}\, .
\en
Note that in deriving Eqs.~(\ref{eq:LSfin},\ref{eq:M})
 we have brought the energy 
denominator of Eq.~(\ref{LS-eqn}) to the non-relativistic form 
$({\bf q}^2-k^2)^{-1}$, eliminating the square roots
 by multiplying the numerator and the denominator in
this equation by the
same algebraic expression and neglecting off-shell terms, which are
exponentially suppressed.

The quantity ${\cal M}_{J'L'M',JLM}(k)$ is related to its counterpart for
spin-zero particles~\cite{Luescher_torus}, according to
\begin{eqnarray}
 {\cal M}_{J'L'M',JLM}(k) = \sum_{m' m \sigma}
                                  {\cal M}_{L'm',L m}(k)\,
 \langle L'm'\,\frac{1}{2}\,\sigma |J'M'\rangle \langle
                                  Lm\,\frac{1}{2}\,\sigma|JM\rangle\, ,
\label{M_jlm}
\end{eqnarray}
where~\cite{Luescher_torus}
\begin{equation}
{\cal M}_{L'm',Lm}(k)=\frac{(-)^{L'}}{\pi^{3/2}}\sum_{j=|L-L'|}^{L+L'}
\sum_{s=-j}^j\frac{i^j}{\kappa^{j+1}}~Z_{js}(1;\kappa^2)
C_{L'm',js,Lm}\, ,
\end{equation}
with $\kappa=kL/(2\pi)$ and
\begin{equation}
Z_{lm}(t;\kappa^2)=\sum_{{\mathbf n}\in \mathbb{Z}^3}\frac{|{\mathbf
    n}|^l\,Y_{lm}(\hat{\mathbf n})}{({\mathbf n}^2-\kappa^2)^t}\, .
   \label{zeta}
\end{equation}
The coefficients $C_{L'm',js,Lm}$ are expressed through the Clebsch-Gordan 
coefficients
\begin{eqnarray}
C_{L'm',js,Lm} = i^{L'-j+L} \sqrt{\frac{(2L'+1)(2j+1)}{(2L+1)}}
\,\langle L'0\,j0|L0\rangle\langle L'm'\,js|Lm\rangle\, .
\end{eqnarray}
Due to Eq.~(\ref{M_jlm}) and the symmetry properties of
 ${\cal M}_{L'm',Lm}$ (see \cite{Luescher_torus}), one finds that
\begin{equation}
 {\cal M}_{JLM,J'L'M'}={\cal M}_{J'L'M',JLM}\, .
\end{equation}
The quantity $t_{A'A}(k,k;z(k))$ defined by the finite-volume LS 
equation~(\ref{eq:LSfin}) develops poles at the momenta where the determinant
of this linear system of equations vanishes. Expressing $h_{JL}(k,k)$ through
the infinite-volume phase shift according to 
Eq.~(\ref{H-by-phase}), we finally obtain the L\"uscher formula for
pion-nucleon scattering
\begin{eqnarray}\label{det-eqn}
 \det \left[ \tan \delta_{J' L'}(k) {\cal M}_{J'L'M',JLM}(k)
                - \delta_{J'J} \delta_{L'L} \delta_{M'M}
\right]=0\, .
\end{eqnarray}
This formula relates the location of the energy eigenvalues
of the pion-nucleon system, placed in a finite box, to the infinite volume
partial-wave phase shifts.

In analogy to Ref.~\cite{Luescher_torus}, it is possible to use the cubic 
symmetry on the lattice in order to achieve the partial block-diagonalization
of the matrix ${\cal M}_{J'L'M',JLM}(k)$. Such a reduction will be considered
in the next section.

\section{Reduction of L\"uscher's formula}
\label{sec:cubic}

In the infinite volume, the basis vectors of the irreducible representation
$D^J$ of the rotation group, corresponding to the total momentum $J$,
are given by $|JLM\rangle=\sum_{ms}|Lm\,\frac{1}{2}\,s\rangle
\langle Lm\,\frac{1}{2}\,s|JM\rangle$.  
Here, $J=\frac{1}{2}\,,\frac{3}{2},\ldots$,
$M=-J,\ldots J$ and $L=J\pm\frac{1}{2}$. The vectors $|JLM\rangle$ are
 also  parity 
eigenvectors with the eigenvalue $P=(-)^L$. Below, we shall use a notation
where the parity is explicitly indicated $|JLM\rangle=|JM\rangle^\pm$.

In a finite volume, for the case of particles with the half-integer spin,
the symmetry breaks down to $^2O\otimes S_2$, where 
$^2O$ denotes the double cover of the cubic group containing 48 elements, no
reflections included (see, e.g.~\cite{Johnson})
and $S_2$ is the discrete group of space inversions. The irreducible 
representations of this group are $G_1^\pm$, $G_2^\pm$ and $H^\pm$ (see 
appendix~\ref{app:basisvectors} for  details). The linear space spanned
by the vectors $|JM\rangle^\pm$ forms a basis of a reducible representation
of the group $^2O\otimes S_2$. We denote the basis vectors, corresponding 
to the irreducible representations, as 
\begin{eqnarray}
 \vert \Gamma,\alpha,J, n \rangle^\pm\, ,
\quad\quad \alpha=1,\ldots {\rm dim}\,\Gamma\, ,
\quad\quad n=1, \ldots N(\Gamma,J)\, .
\end{eqnarray}
Here, $\Gamma=G_1,G_2~\mbox{or}~H$, $N(\Gamma,J)$ denotes the multiplicity
of the irreducible representation $\Gamma^\pm$ in $D^J$ and the index $\alpha$
labels the vectors of a particular irreducible representation.

The basis vectors  $|\Gamma,\alpha,J,n \rangle^\pm$
can be expressed through linear combinations
of  $|JM\rangle^\pm$
\eq
|\Gamma,\alpha,J,n \rangle^\pm
=\sum_M c^{\,\Gamma n\alpha}_{JLM}|JM\rangle^\pm\, .
\en
The matrix elements of the operator ${\cal M}(k)$ in the new basis are given
by
\eq
^\pm\langle {\Gamma'},\alpha',J',n' |{\cal M}(k) | \Gamma,\alpha,J,n \rangle^\pm = \sum_{M'M}(c^{\,\Gamma' n'\alpha'}_{J'L'M'})^* \,
c^{\,\Gamma n\alpha}_{JLM}\,\,{\cal M}_{J'L'M',JLM}(k)\, .
\en
According to Schur's lemma, the operator  ${\cal M}(k)$ is partially 
diagonalized in the new basis
\eq\label{eq:Schur}
^\pm \langle {\Gamma'},\alpha',J', n' | {\cal M}(k)| \Gamma,\alpha,J,n\rangle^\pm =
          \delta_{\Gamma' \Gamma}  \delta_{\alpha' \alpha} 
[{\cal M}^\Gamma_\pm(k)]_{Jn,J'n'}
\en
and equation (\ref{det-eqn}) is rewritten as
\begin{eqnarray}\label{det-eqn1}
\prod_{L'=J'\pm\frac{1}{2}}\prod_\Gamma \det \biggl( \tan \delta_{J'L'}(k) [{\cal M}^\Gamma_\pm(k)]_{J'n',Jn}
                - \delta_{J'J} \delta_{n'n} \biggr)=0\, ,
\end{eqnarray}
where the $+/-$ sign in ${\cal M}^\Gamma_\pm$ corresponds to  even/odd $L'$.

In table~\ref{tab:M_nl} we list the matrix elements
 $[{\cal M}^\Gamma_\pm(k)]_{J'n',Jn}$ for
$J',J<\frac{9}{2}$. Since the multiplicity $N(\Gamma,J)=1$ for $J<\frac{9}{2}$,
the indices $n',n$ can be omitted  in this table. The entries of the table
are expressed through the following quantities~\cite{Luescher_torus}
\begin{equation}
 {\cal W}_{lm}=\left( \pi^{3/2}(2l+1)^{1/2} \kappa^{l+1} \right)^{-1} 
Z_{lm}(1;\kappa^2)\, .
\end{equation}
The construction of the basis vectors in case of arbitrary $J$ is considered 
in appendix~\ref{app:basisvectors}.

\begin{table}[t]

\renewcommand{\arraystretch}{1.4}

\begin{center}
\begin{tabular}{|c|cc|c|}\hline
$\Gamma^\pm$  &  $J$    &   $J'$    & ${\cal M}^\Gamma_\pm(k)$      \\\hline
$G_1^\pm$     &  $1/2$  &  $1/2$    & ${\cal W}_{00}$               \\
$G_1^\pm$     &  $1/2$  &  $7/2$    & $\mp \frac{4\sqrt{21}}{7}\,\,{\cal W}_{40}$   \\
$G_1^\pm$     &  $7/2$  &  $7/2$    & ${\cal W}_{00}
+\frac{18}{11}\,{\cal W}_{40}+\frac{100}{33}\,{\cal W}_{60}$   \\
$G_2^\pm$     &  $5/2$  &  $5/2$    & ${\cal W}_{00}-\frac{12}{7}\,{\cal W}_{40}$ \\
$G_2^\pm$     &  $5/2$  &  $7/2$    & $\pm\frac{60\sqrt{3}}{77}\,{\cal W}_{40}
\mp\frac{40\sqrt{3}}{11}\,{\cal W}_{60}$ \\
$G_2^\pm$     &  $7/2$  &  $7/2$    & ${\cal W}_{00}
-\frac{162}{77}\,{\cal W}_{40}+\frac{20}{11}\,{\cal W}_{60}$ \\
$H^\pm$     &  $3/2$  &  $3/2$    & ${\cal W}_{00}$ \\
$H^\pm$     &  $3/2$  &  $5/2$    & $\mp\frac{6\sqrt{6}}{7}\,{\cal W}_{40}$   \\
$H^\pm$     &  $5/2$  &  $5/2$    & ${\cal W}_{00}+\frac{6}{7}\,{\cal W}_{40}$ \\
$H^\pm$     &  $3/2$  &  $7/2$    & $\frac{2\sqrt{30}}{7}\,{\cal W}_{40}$   \\
$H^\pm$     &  $5/2$  &  $7/2$    & $
\mp\frac{36\sqrt{5}}{77}\,{\cal W}_{40}\mp\frac{20\sqrt{5}}{11}\,{\cal W}_{60}$   \\
$H^\pm$     &  $7/2$  &  $7/2$    & ${\cal W}_{00}+\frac{18}{77}~{\cal W}_{40}-\frac{80}{33}~{\cal W}_{60}$   \\
\hline
\end{tabular}
\end{center}
\caption{Non-vanishing matrix elements $[{\cal M}^\Gamma_\pm(k)]_{Jn,J'n'}$ for 
$J$, $J' < 9/2$ and $n=n'=1$. The matrix is symmetric under $J'n'\leftrightarrow Jn$.}
\label{tab:M_nl}
\end{table}

\section{Probability distribution}
\label{sec:probability}

As mentioned in the introduction, in the vicinity of a narrow resonance
the finite-volume energy levels of a two-particle system exhibit the peculiar
behavior known as the avoided level crossing. Such a behavior, which is
predicted by L\"uscher's formula, is schematically shown in 
Fig.~\ref{fig:sceme-of-analysis}. In this figure, we plot the relative
momentum $p$, which is related to the CM energy $E$ as 
$E=(m_N^2+p^2)^{1/2}+(M_\pi^2+p^2)^{1/2}$, vs the box size $L$. The plateaus
correspond to the resonance energy and the resonance width is determined
by the minimal distance between the curves.

It was, however, also mentioned above that for most physically interesting
strong resonances the avoided level crossing is almost completely washed
out from the spectrum due to the large width of a resonance. In this section,
we describe a method that can be used to visualize 
the extraction of the resonance parameters
from the two-particle spectrum even in this case. What is important is that the
method does not contain any prior theoretical bias (e.g. does not use the
resonance parameterization of the infinite-volume scattering phase 
as an input).

Assume now that the volume-dependent two-particle spectrum  is
measured on the lattice.
 The probability distribution $W(p)$ is constructed according to the
following prescriptions:

\begin{figure}[t]
\begin{center}
\includegraphics[width=10.cm]{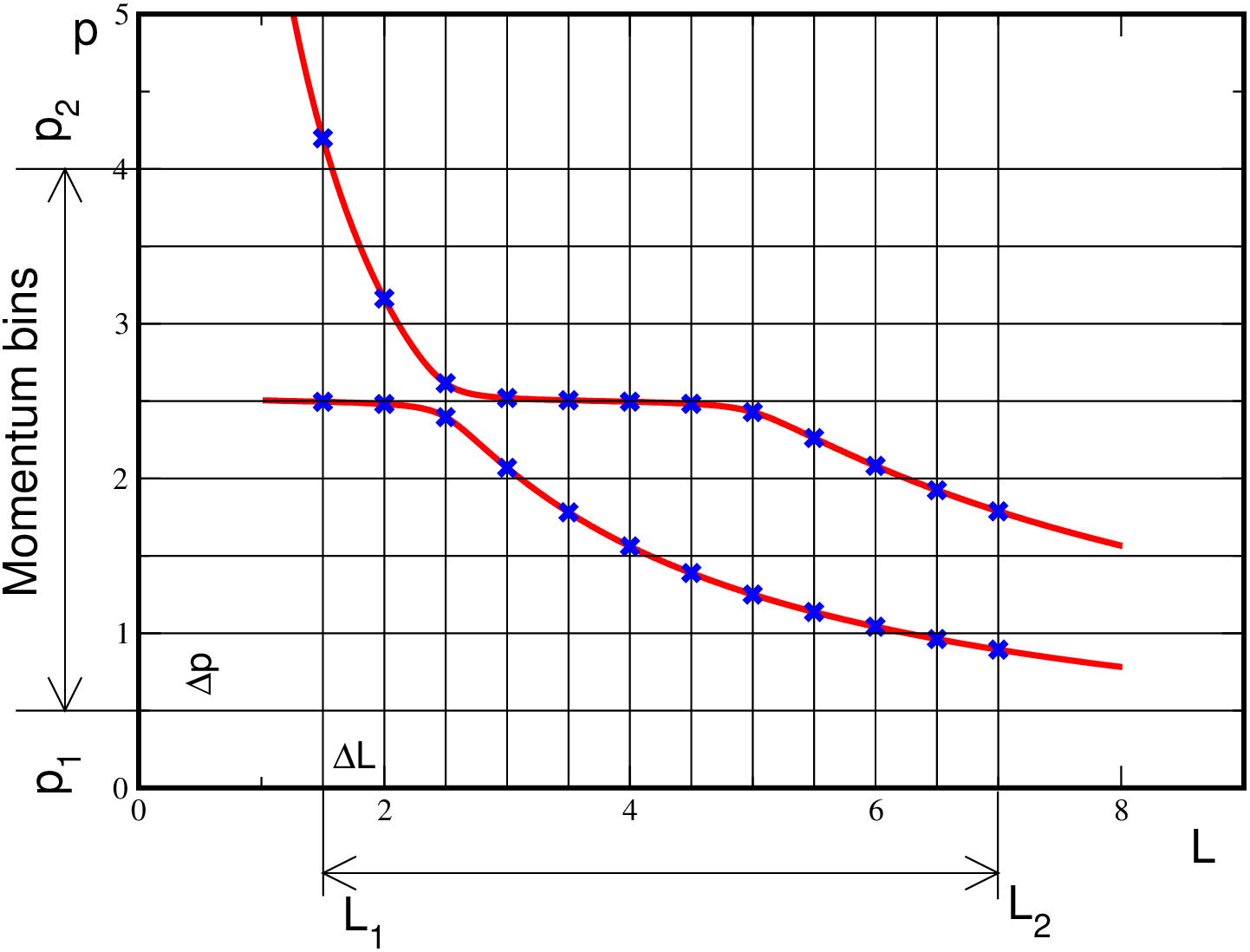}
\end{center}
\caption{Schematic plot describing the construction of the probability 
distribution from the energy levels.}
\label{fig:sceme-of-analysis}
\end{figure}

\begin{itemize}
\item[i)]
Choose the first $N$ energy levels (e.g., $N=2$ in Fig.~\ref{fig:sceme-of-analysis}); choose the interval $L\in[L_1,L_2]$
and slice this interval into equal parts with length $\Delta L$. For
each value of $L=L_1$, $L=L+\Delta L$, etc determine
$p_n(L),~n=1\cdots N$.

\item[ii)]
Choose a corresponding momentum interval $p\in [p_1,p_2]$ and introduce equal-size
momentum bins with length $\Delta p$.

\item[iii)]
Count, how many times the eigenvalue $p_n(L),~n=1\cdots N$ is contained in a particular bin, 
if $L$ runs from $L_1$ to $L_2$.
This number gives the unnormalized probability distribution for the momentum 
bin chosen. 
Normalizing this distribution
in the interval $[p_1,p_2]$ yields finally the probability distribution 
$W(p)$ we are looking for. The normalization condition is given by
\begin{equation}
\sum_{k=0}^{M_P} W(p_1+k\Delta p)\Delta p=1\, ,
\quad M_P=\frac{p_2-p_1}{\Delta  p}-1\, .
\end{equation}
\end{itemize}

It is clear that in case of a pronounced avoided level crossing, the
probability distribution $W(p)$ 
must be strongly peaked around the resonance energy. 
The exact shape of $W(p)$ can be predicted on the basis of L\"uscher's
formula. To this end, note that, in the limit of infinitesimally small
$L-$ and $p-\,$bins, $W(p)$ is given by
\begin{eqnarray}\label{eq:pnL}
 W(p)=C \sum_{n=1}^{N} \frac{1}{p'_n(L)}\, ,
\end{eqnarray}
where $C$ is a normalization constant.

For simplicity, we
consider here the scattering in the partial wave with $L=1,~J=\frac{3}{2}$ and
neglect the (small) mixing to higher partial waves. Using table~\ref{tab:M_nl}
and Eq.~(\ref{det-eqn1}), the
relation between the finite-volume energy spectrum and the phase shift
$\delta(p)$ takes the form\footnote{To ease notation, we do not attach indices $L,J$
to this phase shift.}
 \begin{equation}
 \delta(p)=-\phi(\kappa)+\pi n,\quad\quad
\phi(\kappa)=-\arctan\frac{\pi^{3/2}\kappa}{Z_{00}(1;\kappa^2)}\, ,\quad\quad
\kappa=\frac{pL}{2\pi}\, , \label{three-dim-luscher}
\end{equation}
where the integer $n$ labels the energy levels $p_n(L)$, which are the
solutions of the above equation.

Differentiating now Eq.~(\ref{three-dim-luscher}) with respect to $L$ and
substituting into Eq.~(\ref{eq:pnL}), we obtain
\begin{equation}\label{eq:W}
 W(p)=C\sum_{n=1}^{N}\left( \frac{L_n(p)}{p} 
+ \frac{2\pi \delta'(p)}{p \phi'(\kappa_n(p))}\right)\, .
\end{equation}
where $\kappa_n(p)$ and $L_n(p)$ are the solutions of 
Eq.~(\ref{three-dim-luscher}) for a given $p$. It is seen that $W(p)$ defined
by Eq.~(\ref{eq:W}) is closely related to the so-called ``density of
states in a finite volume,'' see e.g. Ref.~\cite{Lin}.

In the vicinity of the resonance, $\delta'(p)$ is strongly
peaked. Substituting  a Breit-Wigner parameterization for $\delta(p)$ and
assuming that all other factors smoothly depend on the momentum $p$, we 
may verify that in the vicinity of the resonance the function $W(p)$ follows the
Breit-Wigner form for the scattering cross section, with the same 
width\footnote{Note that 
the location of the maximum in the probability distribution
does not, in general, coincide either with the real part of the pole position
in the amplitude, or with the solution of the equation $\delta(p_R)=\pi/2$.
These three quantities agree only in the limit of the infinitely 
narrow resonance.}. 

A useful parameterization of $W(p)$ can be obtained by using the following
approximation of the function $\phi(\kappa)$, which is valid in a large 
interval of arguments~\cite{Luescher_rho}
\eq
\phi(\kappa)=\pi c \kappa^2\, , \qquad c\simeq 1\, .
\en
Solving L\"uscher's equation, we obtain
\eq
L_n(p)=\frac{1}{p}\sqrt{4\pi(\pi n-\delta(p))}
\en
and
\eq
W(p)= \frac{C}{p}\,\sum_{n=1}^N\biggl(\frac{\sqrt{4\pi(\pi n-\delta(p))}}{p}
+\frac{2\pi\delta'(p)}{\sqrt{4\pi(\pi n-\delta(p))}}\biggr)\, .
\en
In order to suppress the (large) background, related to the free motion
of the $\pi N$ pair, we consider in the following the so-called subtracted
probability distribution $W(p)-W_0(p)$, where $W_0(p)$ is determined
from Eq.~(\ref{eq:W}) with $\delta(p)=0$ and $L_n(p)$ corresponding to the
free energy levels.

It is interesting to consider the infinite-volume limit of the
 probability distribution. Note that in this limit the number of
energy levels per fixed momentum bin goes to infinity. Consequently, the
number of levels $N$ in Eq.~(\ref{three-dim-luscher}) can be chosen very
large. In this case, in the expression for the quantity $L_n(p)$ which is
determined through the solution of L\"uscher's equation
\begin{equation}
 L_n(p)=\frac{2\pi}{p} \phi^{-1}\left(\pi n - \delta(p)\right)\, ,
\end{equation}
the phase shift obeys the inequality $\delta(p)\ll\pi n$ for the large
majority of terms in the sum over energy levels. Expanding in a Taylor series,
we get
\begin{equation}
 L_n(p)= \frac{2\pi}{p}\,\bar{\kappa}_n
-\frac{2\pi}{p}\delta(p)\frac{1}{\phi'(\bar{\kappa}_n)}+O(\delta^2)~,
\end{equation}
with $\bar{\kappa}_n$ is the solution of the equation 
$\phi(\bar{\kappa}_n)=\pi n$.
Using, in addition, $\kappa_n=\bar{\kappa}_n+O(\delta)$, the unnormalized
probability distribution for $N\to \infty$ takes the form
\begin{equation}
 C^{-1}W(p)=  \frac{2\pi}{p^2} \sum_{n=1}^N \bar{\kappa}_n 
                               + \frac{2\pi}{p} \sum_{n=1}^N \frac{1}{\phi'(\bar{\kappa}_n)}
                                 \left( \frac{\delta(p)}{p} - \delta'(p) \right) 
                                  + O(\delta^2)~.
\end{equation}
The first term exactly coincides with the free background. Subtracting
this background and taking into account the fact that the quantity
$\sum_{n=1}^N (\phi'(\bar{\kappa}_n))^{-1}$ does not depend
on the phase $\delta(p)$, we obtain 
\begin{equation}
C^{-1} W(p)-C_0^{-1}W_0(p) \propto \frac{1}{p} \left( \frac{\delta(p)}{p} -
  \delta'(p) \right)\, .
\end{equation}
In other words, in the infinite-volume limit this quantity is determined by the
elastic phase shift alone.

\begin{figure}[t]
\begin{center}
\includegraphics[width=10.cm]{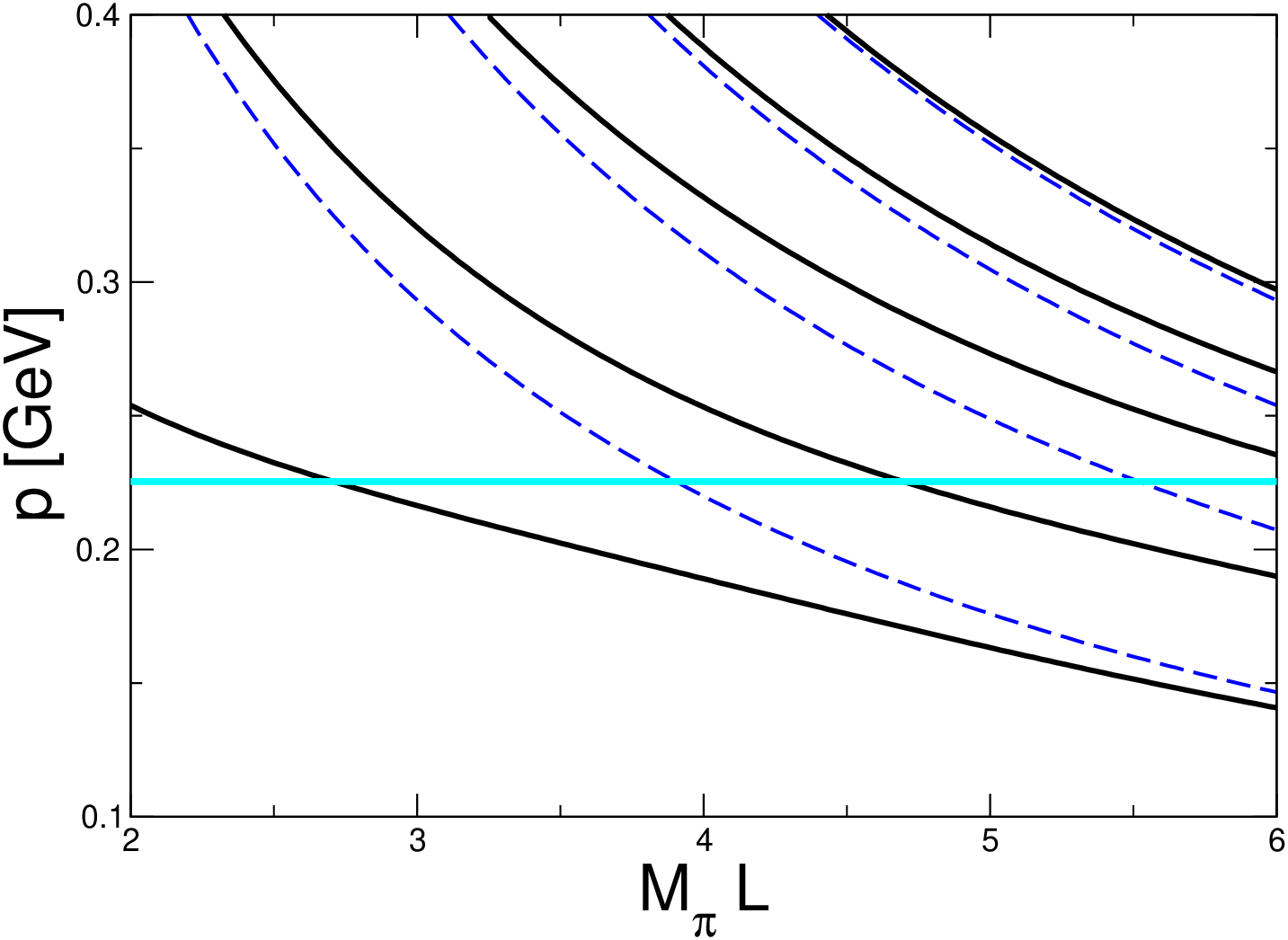}
\end{center}
\caption{Volume dependence of the spectrum of the Hamiltonian (solid lines),
predicted by using L\"uscher's formula with input experimental $P_{33}$ phase 
shift. The center-of-mass momentum $p$ vs the box size $L$ is shown.
For comparison, the free energy levels for the $\pi N$ system are given
(dashed lines). The horizontal line marks the position of the $\Delta$ 
resonance. It is seen that the avoided level crossing is completely washed out.
  }
\label{fig:deltaenergies}
\end{figure}

\section{Analysis of synthetic data}
\label{sec:analysis}

In this section we will implement the 
method discussed in the previous section for analyzing  data. In the 
absence of lattice QCD data we will use  
synthetic data on the spectrum of
the Hamiltonian, which are produced by using
experimentally measured phase shifts~\cite{SAID} in L\"uscher's formula.
If in the future unquenched lattice calculations are performed at the physical
quark masses, the results must agree with the above synthetic data set.

\begin{figure}[t]
\begin{center}
\includegraphics[width=6.cm]{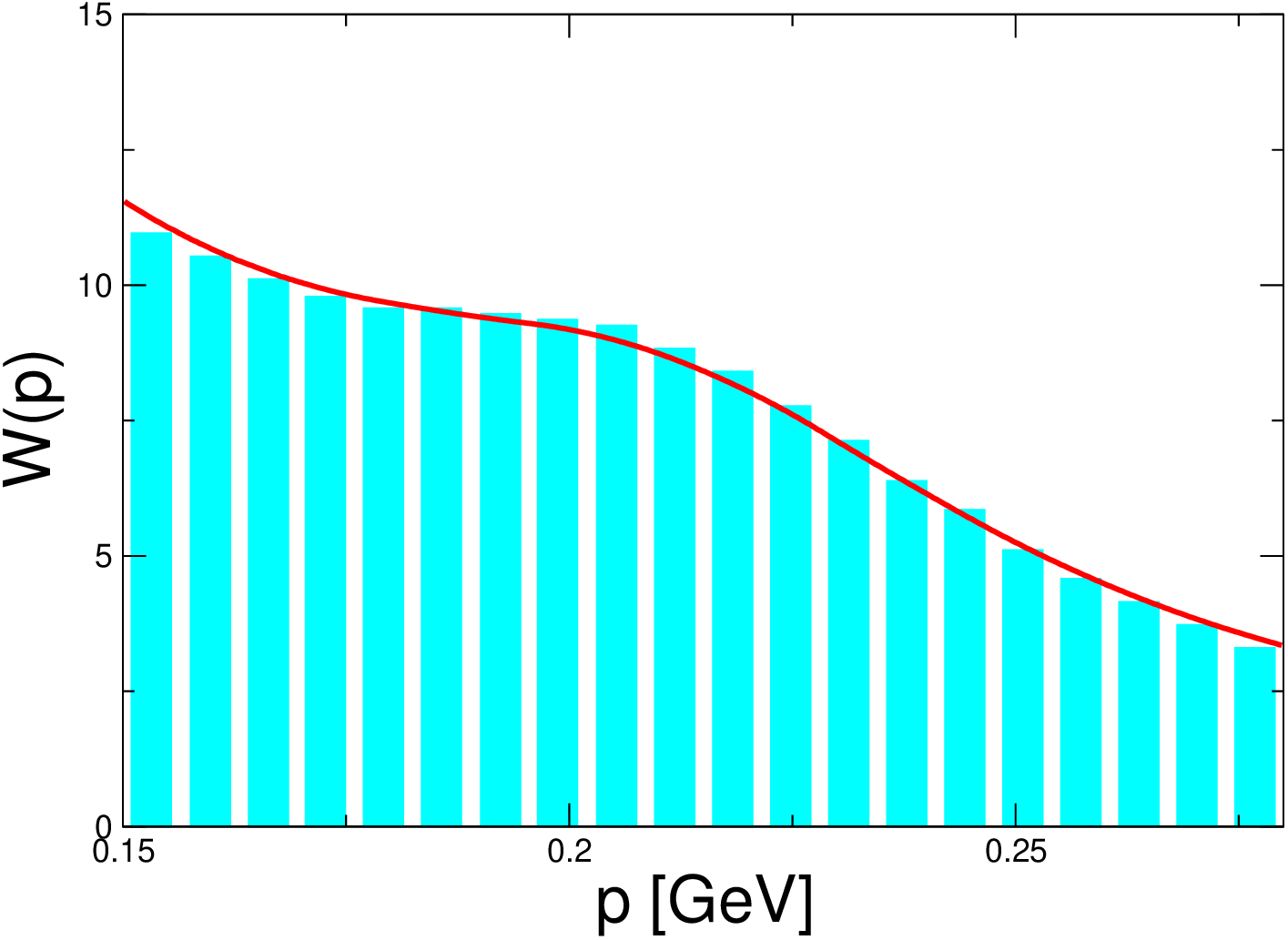}
\hspace*{.6cm}
\includegraphics[width=6.cm]{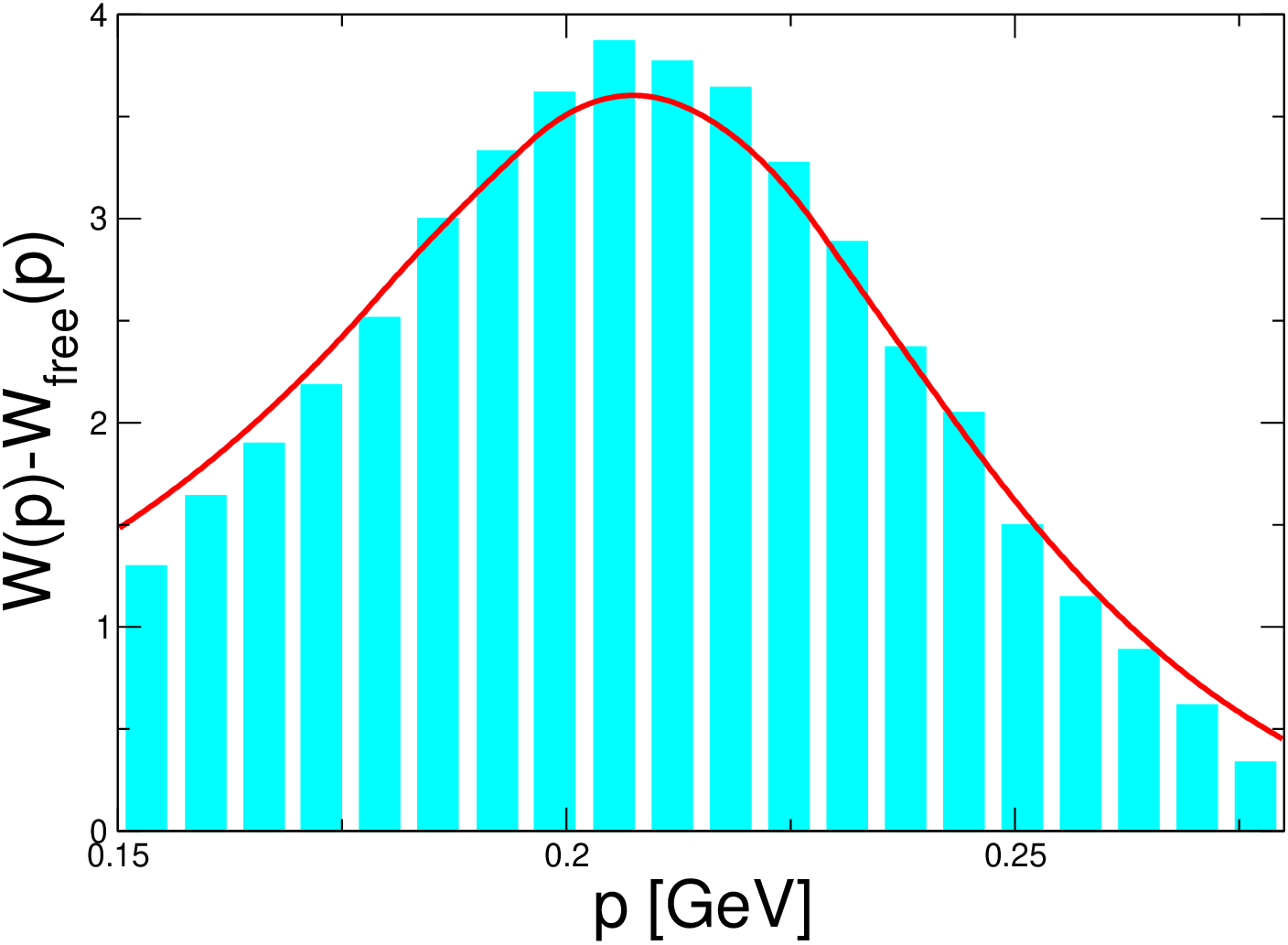}
\end{center}
\caption{Unsubtracted (left panel) and subtracted (right panel) 
probability distributions. Only the lowest energy level has been included
in the analysis ($N=1$). 
The solid lines correspond to the prediction made by using
L\"uscher's formula with an approximation $\phi(\kappa)\simeq \pi \kappa^2$ (see the text
for details). A clear resonance-like structure is observed in the 
subtracted distribution.}
\label{fig:subtraction}
\end{figure}

\begin{figure}[t]
\begin{center}
\includegraphics[width=6.cm]{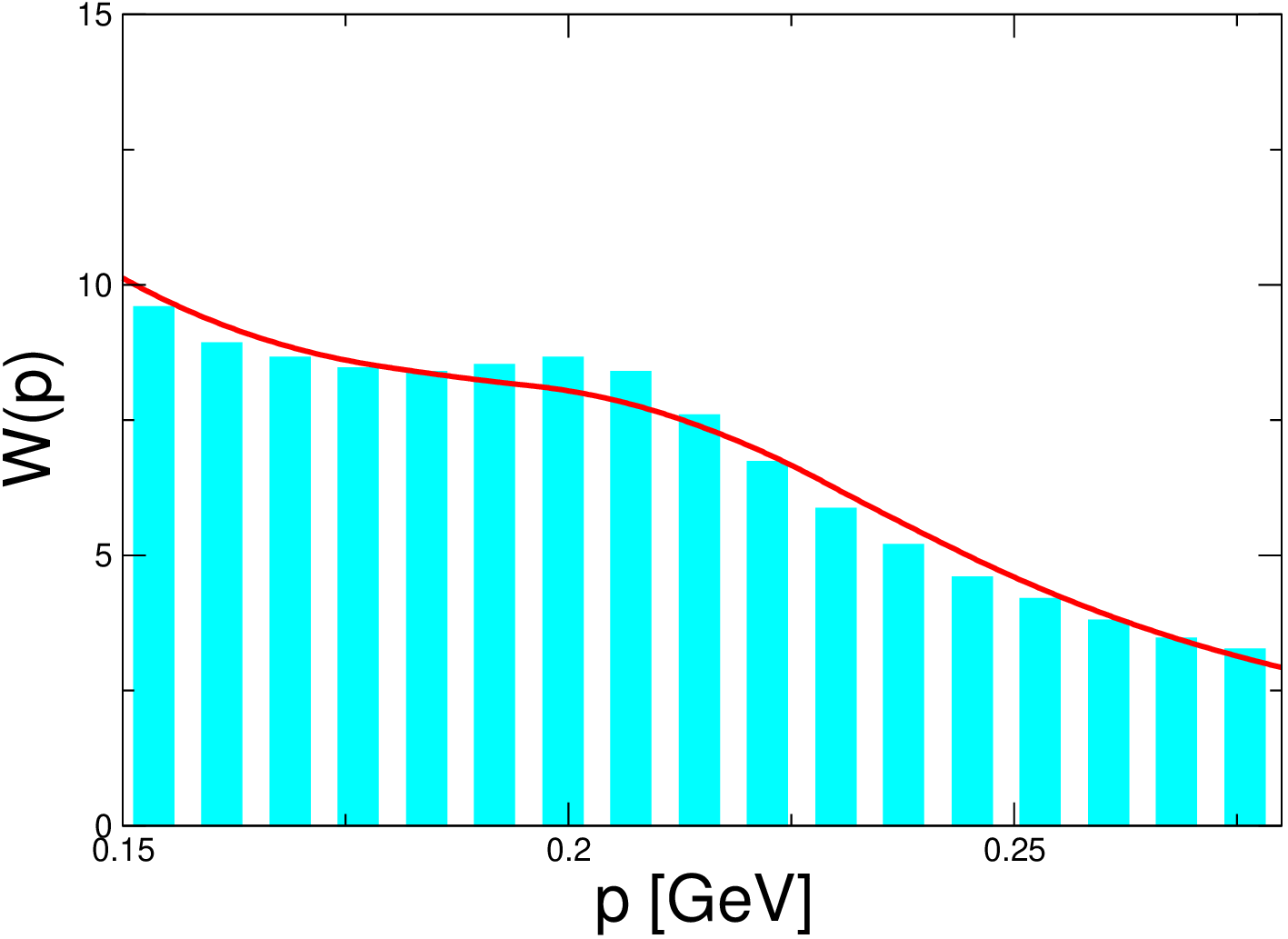}
\hspace*{.6cm}
\includegraphics[width=6.cm]{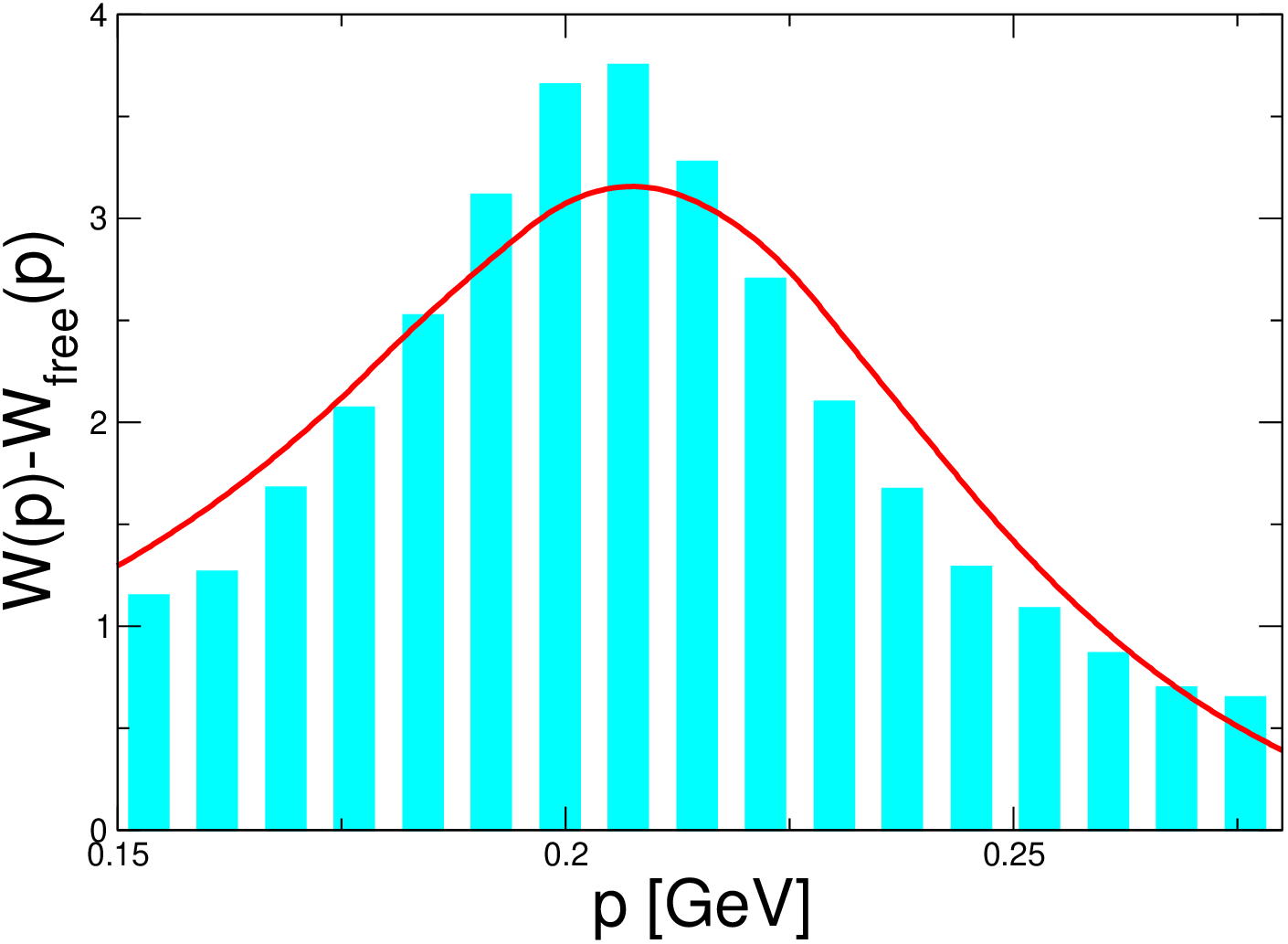}
\end{center}
\caption{The same as in Fig.~\ref{fig:subtraction} but obtained only
from 5 data points corresponding to $M_\pi L=1.4,~2.55,~3.7,~4.85,~6$.
The values of the function $p(L)$ between the data points are obtained using
Spline interpolation. Only the lowest energy level is analyzed ($N=1$).
The interval in the variable $p$ is the same as in Fig.~\ref{fig:subtraction}.
It is seen that 5 data points
do not provide a very good accuracy
over this interval in $L$: the subtracted probability distribution is rather
different from the theoretical prediction made on the basis of L\"uscher's 
formula.}
\label{fig:5points}
\end{figure}

The calculated spectrum, obtained by substituting the
resonant $P_{33}$ partial wave phase shift into L\"uscher's formula,
is shown in Fig.~\ref{fig:deltaenergies}. In this figure, the relative
momentum of the $\pi N$ system $p(L)$, corresponding to the discrete 
energy levels of the Hamiltonian in a finite box, 
is plotted against the box size $L$
(in units of $M_\pi^{-1}$). It is seen that the structure of the energy levels
is smooth: the avoided level crossing has been completely washed out due
to the relatively large width of $\Delta$. However, in the 
same figure we also plot the free energy levels, demonstrating
that in the vicinity of the resonance a continuous
rearrangement of the spectrum takes place.

This rearrangement can be made explicitly visible by performing
the analysis of the energy spectrum,
using the probability distribution method introduced above. Constructing, as
described above,
the unsubtracted probability distribution $W(p)$ from the lowest energy level
 yields the plot 
shown on the left panel of Fig.~\ref{fig:subtraction}. The resonance
is seen as a barely distinguishable shoulder around $p\simeq 0.22~\mbox{GeV}$.
This result, which obviously reflects the washing-out of the
avoided level crossing in Fig.~\ref{fig:deltaenergies}, casts justified
doubts on the feasibility of a clean extraction of the $\Delta$-resonance
parameters from the data. The picture, however, completely changes once the
subtraction of the background due to free $\pi N$ pairs has been 
performed, see the right panel in Fig.~\ref{fig:subtraction}.
The resonance-like structure in the subtracted distribution
is clearly visible, allowing one to finally conclude that
the determination of the resonance parameters from the data is indeed
possible. On both plots the solid curves correspond to the theoretical
prediction made on the basis of L\"uscher's formula. In order to simplify
the numerical calculations, the curves were constructed 
by using the approximation 
$\phi(\kappa)\simeq \pi \kappa^2$, which works very well for all relevant values of 
the variable $\kappa$. 
Since these curves are shown for the demonstrative purposes
only, a better accuracy is not needed here.

\begin{figure}[t]
\begin{center}
\includegraphics[width=6.cm]{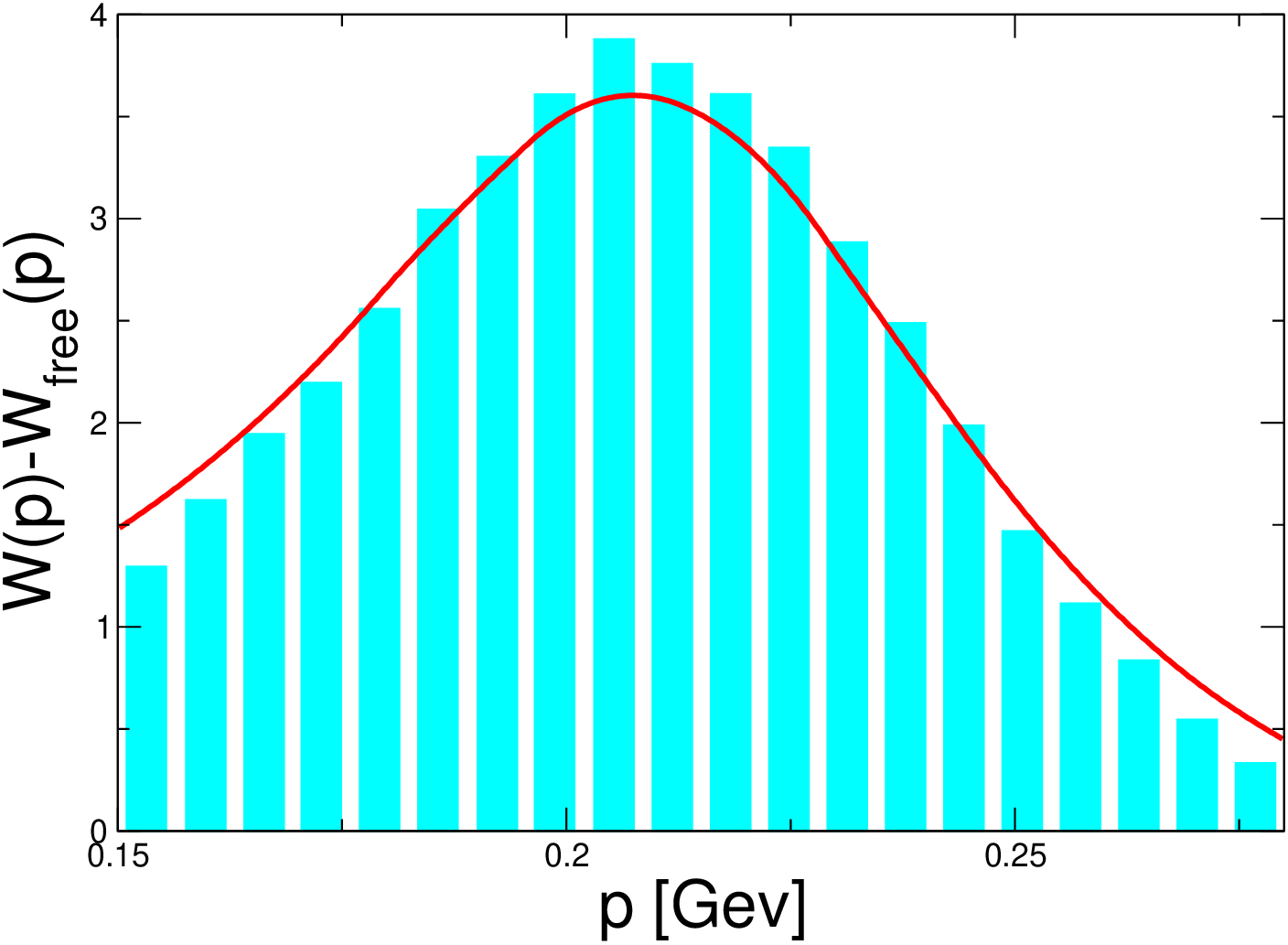}
\hspace*{.6cm}
\includegraphics[width=6.cm]{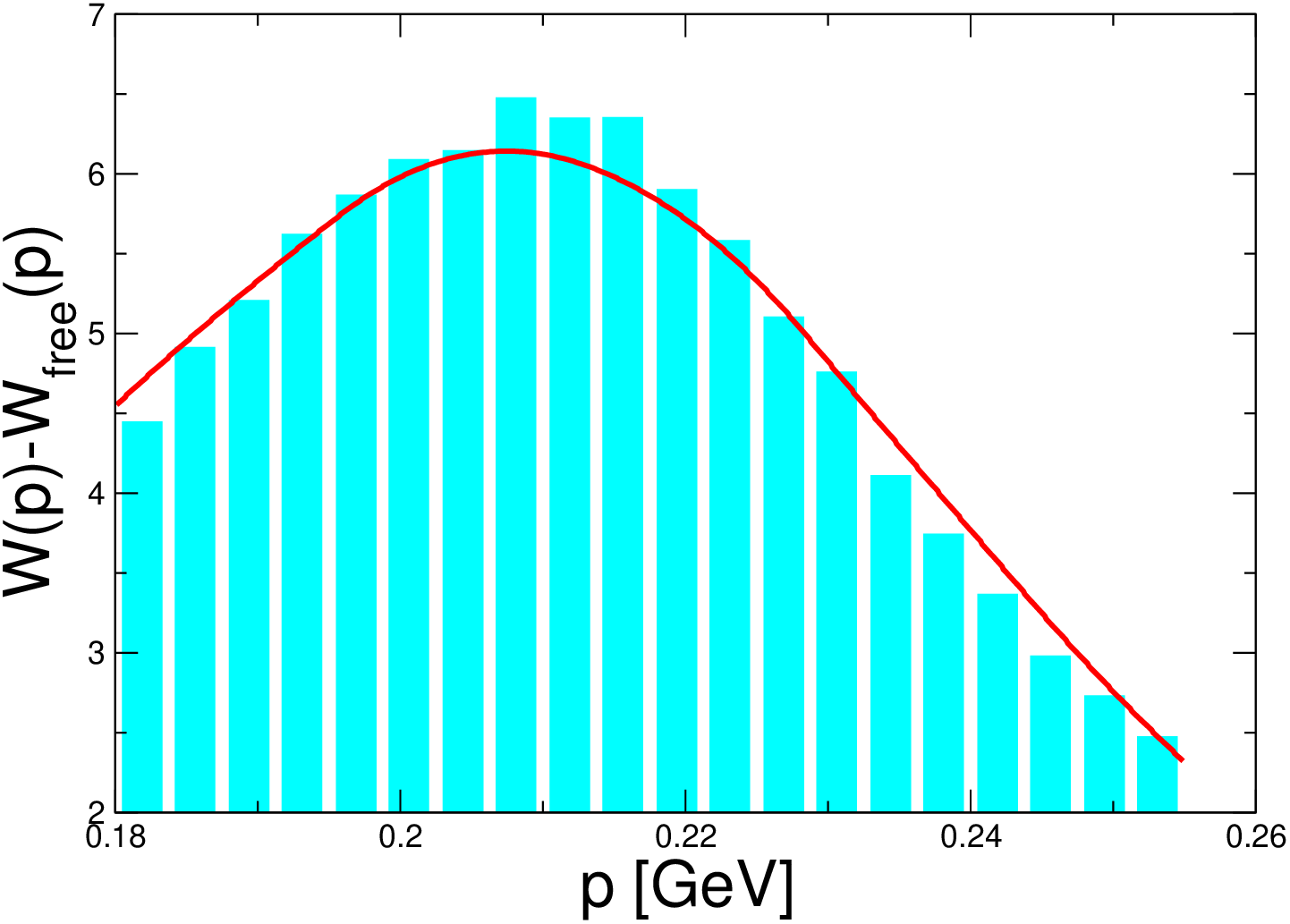}
\end{center}
\caption{The subtracted probability distribution for 10 data points
equally distributed in the interval $M_\pi L=1.4\cdots 6$ (left panel).
The same with 5 data points,  $M_\pi L=1.9\cdots 4.5$
(right panel). In both cases, only the
lowest energy level has been included in the analysis ($N=1$).
The agreement with the theoretical curve, based on L\"uscher's formula,
is much better than for Fig.~\ref{fig:5points} (right panel).
}
\label{fig:improving}
\end{figure}

\begin{figure}[htb]

\vspace*{.2cm}

\begin{center}
\includegraphics[width=6.cm]{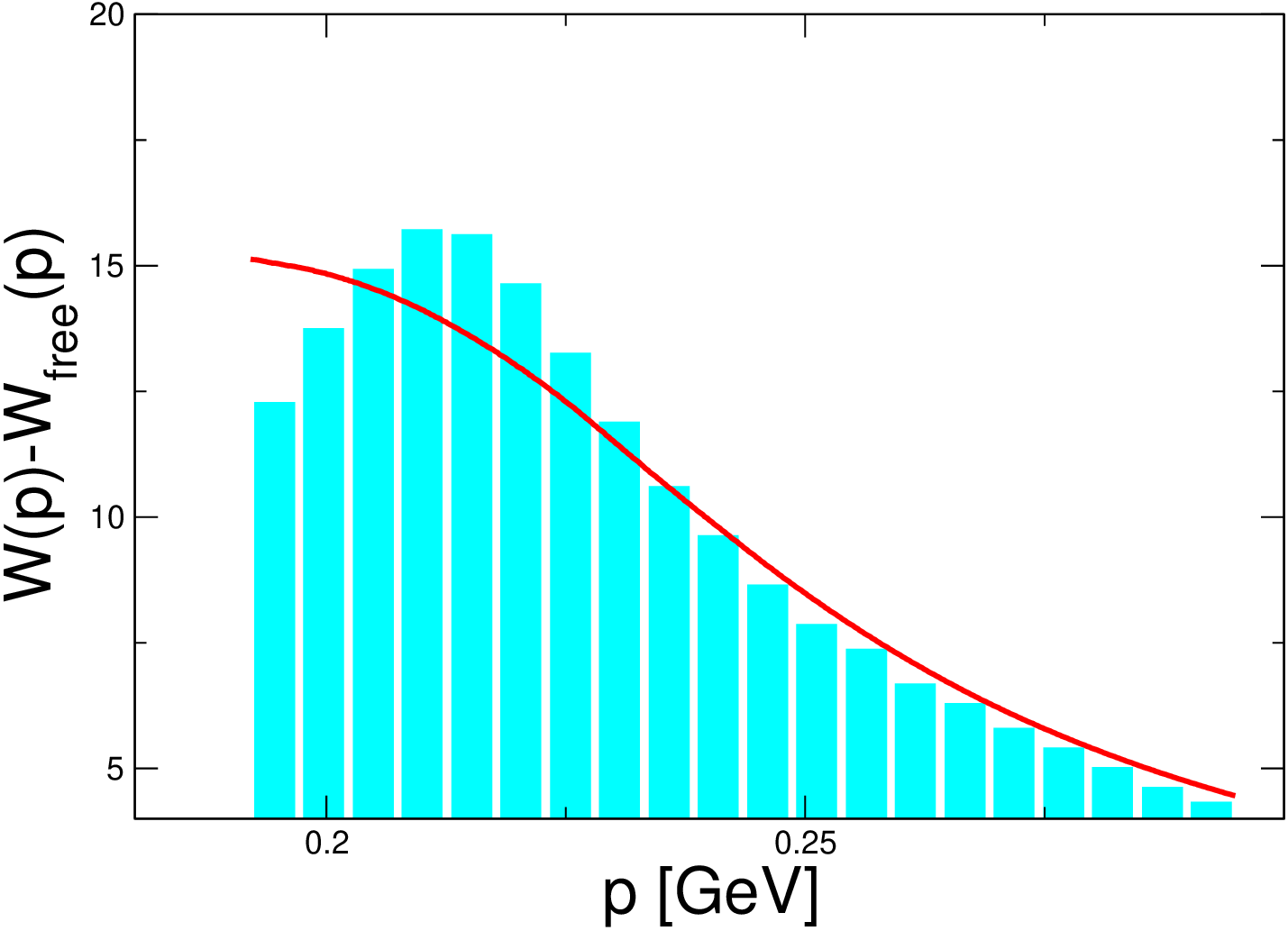}
\end{center}
\caption{A non-uniform choice of the data points 
$M_\pi L=1.4,~1.6,~1.8,~2.8,~3.9$. The momentum interval is chosen
to be $p\in [0.21,0.35]~\mbox{GeV}$. It is immediately seen that the
subtracted probability distribution deviates from the theoretically
predicted behavior in the interval where the data points are sparse
(large volumes or small momenta).}
\label{fig:asym}
\end{figure}

Up to now, we have dealt with the exact solution of L\"uscher's equation for 
the energy spectrum that on the lattice corresponds to measuring this spectrum 
at infinite accuracy and at all values of $L$. The situation in real 
calculations is different. Here one expects to get at most
a few data points for 
different volumes. Our next aim is to mimic this situation in the calculations
with the synthetic data and check whether the extraction of the resonance
parameters is still possible. The probability distribution method, 
which we are using, is equivalent to L\"uscher's approach and 
provides just a nice tool to visualize the final result.

In order to achieve the goal formulated above, we first perform the analysis
in the same momentum interval
as in Fig.~\ref{fig:subtraction}, but using  5 uniformly distributed
data points, located at $M_\pi L=1.4,~2.55,~3.7,~4.85,~6$. 
The values of $p(L)$  between these data points were reconstructed 
by using the interpolation procedure with cubic splines. The result
for the unsubtracted and subtracted probability distributions is shown
in Fig.~\ref{fig:5points}. It is clear that providing only
5 data points at this quite large interval of the variable $L$ does not 
ensure a very high accuracy: the shape of the resonance comes out  distorted. 
Moreover, one might expect that if even less data points are included in the 
analysis (see, e.g.~\cite{Aoki:2007rd}), 
the determined resonance parameters will include large systematic
uncertainty which is very hard to control.

There can be two possible ways out. One may try to gradually increase 
the number of data points, or one may try to reduce the size of the
momentum and volume interval just to an immediate proximity of the resonance.
Both possibilities have been tried, as shown in Fig.~\ref{fig:improving}.
The left panel of this figure corresponds to using 10 data points
instead of 5 in the same interval, whereas the right panel corresponds to reducing
the interval to $M_\pi L\in [1.9,4.5]$,
respectively. In both cases one observes a clear improvement as compared
to the case shown in Fig.~\ref{fig:5points}. Note also that the data
points should not not be uniformly distributed in $L$, but
 have indeed to be concentrated on both sides of the resonance. Otherwise,
one may arrive to the picture shown in Fig.~\ref{fig:asym}, where we 
show the probability distribution, obtained from the following data points:
$M_\pi L=1.4,~1.6,~1.8,~2.8,~3.9$. It is seen that the probability 
distribution significantly deviates
from the exact theoretical prediction in that part of the interval, where
the data points are sparse.

\begin{figure}[t]

\vspace*{.6cm}

\begin{center}
\includegraphics[width=6.cm]{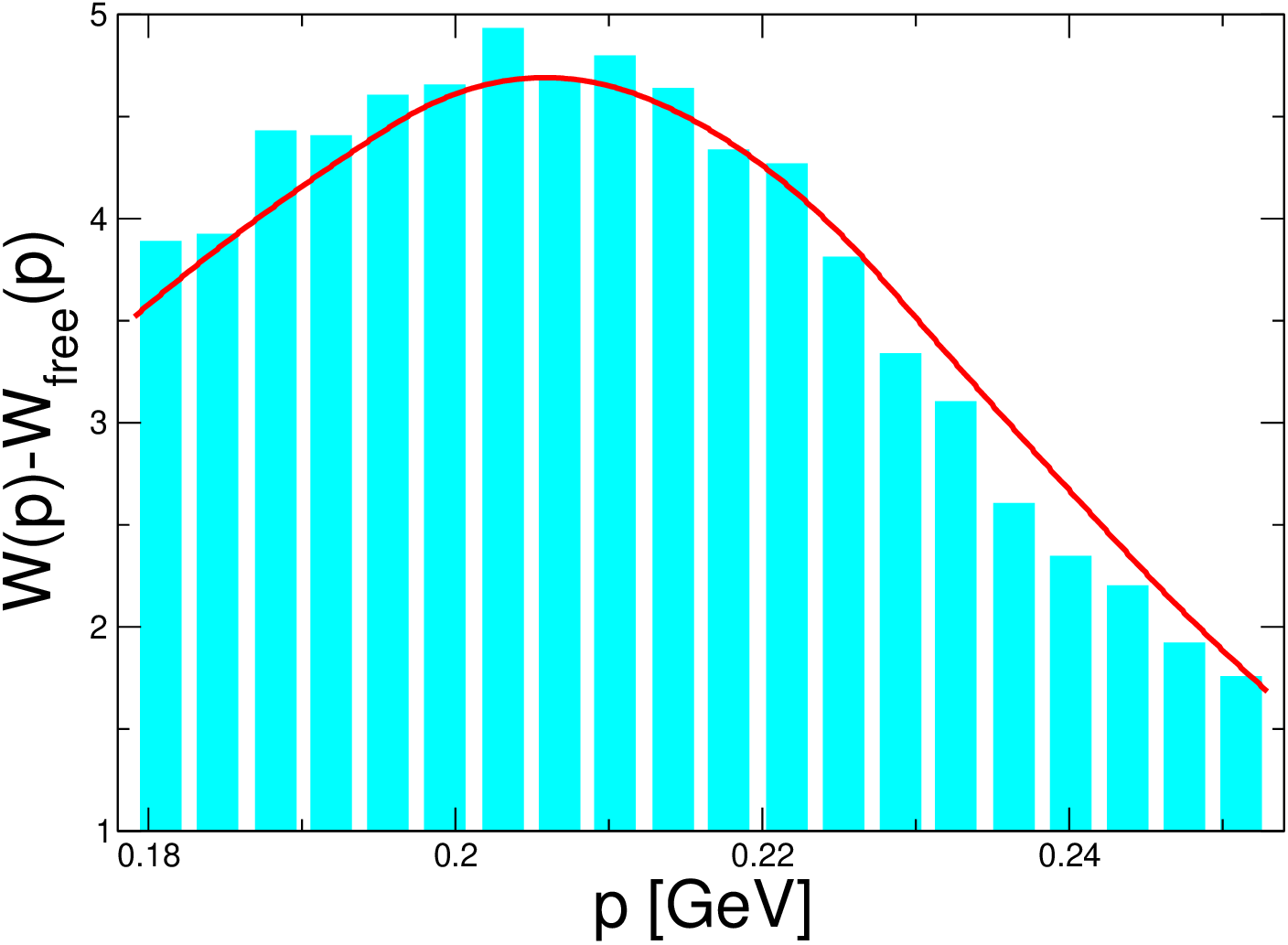}
\end{center}
\caption{The result of the analysis of the lowest two energy levels
($N=2$) with 10 data points,  $M_\pi L=2\cdots 6.5$.
The resonance structure reasonably reproduces
 the theoretical prediction on the basis of L\"uscher's formula.}
\label{fig:2levels}
\end{figure}

Finally, we have applied the method of probability distributions to the 
simultaneous analysis of the first two energy levels. The figure~\ref{fig:2levels}
contains the information about 10 data points, uniformly distributed
in the interval $M_\pi L\in[2,6.5]$. The resulting resonance shape is again
in good agreement with the theoretical prediction, made on the basis of
L\"uscher's formula. Note that the resonance parameters extracted from the
analysis of the different energy levels must of course coincide.
In case the data from the excited levels are also available, checking
the stability of the resonance parameters might enable one to verify
{\it a posteriori},
whether the volumes, used in the calculation, are large enough to justify
the application of L\"uscher's approach.

Last but not least, the lattice data on the measured spectrum
always come with errors.
 Our method provides an easy tool
to render the analysis transparent in this case as well. The procedure is 
described below.

\begin{figure}[t]
\begin{center}
\includegraphics[width=5.0cm]{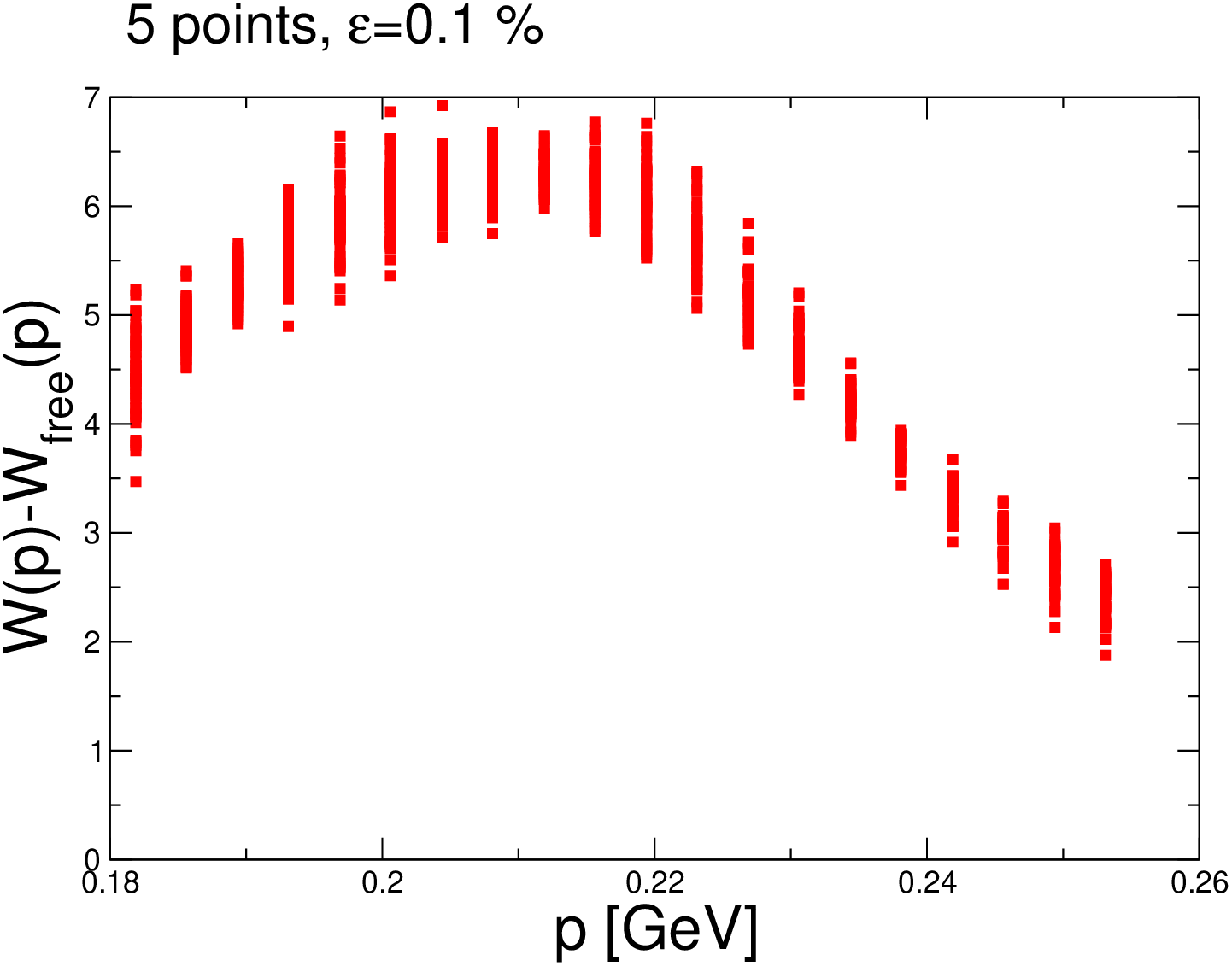}\hspace*{.2cm}
\includegraphics[width=5.0cm]{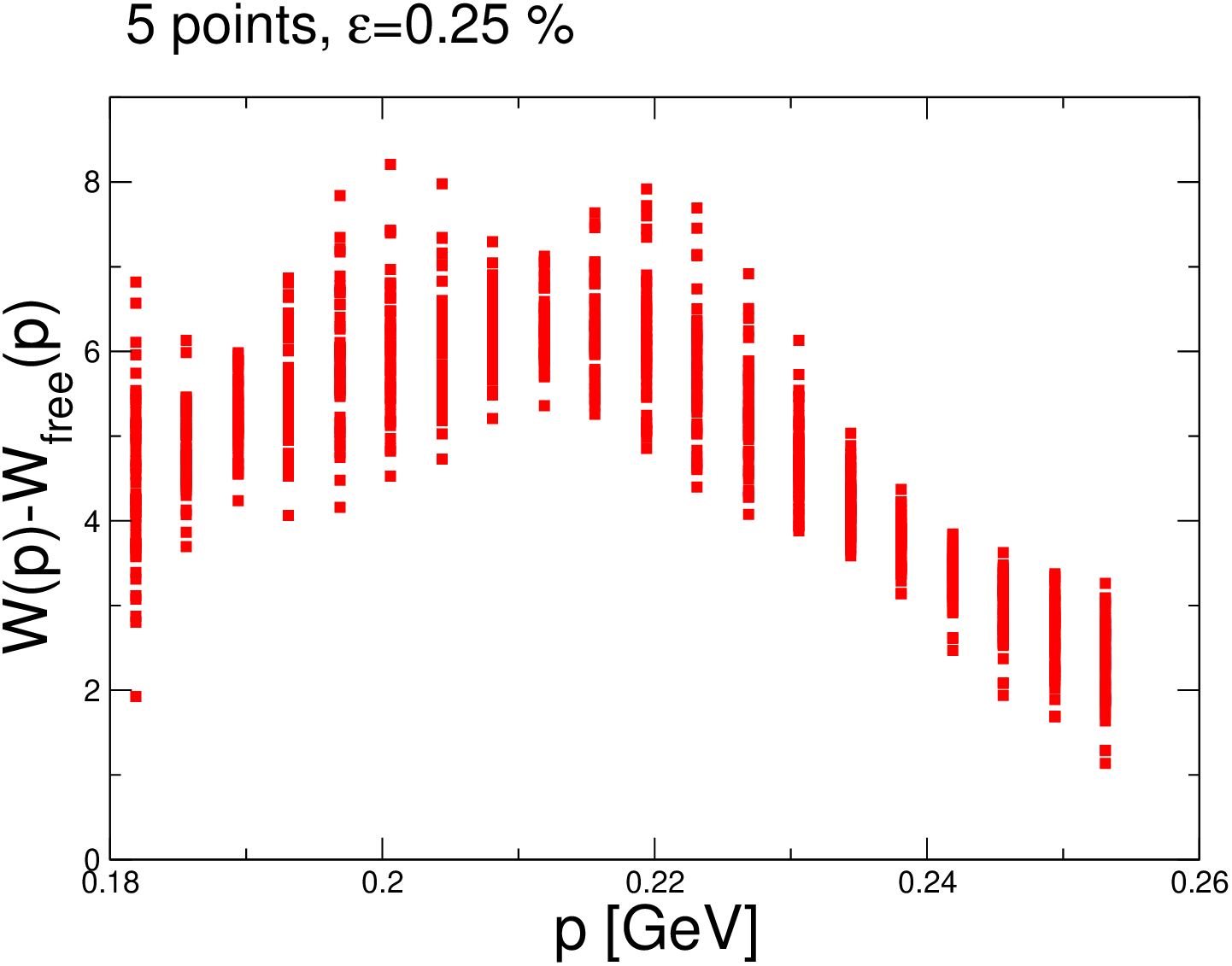}\hspace*{.2cm}
\includegraphics[width=5.0cm]{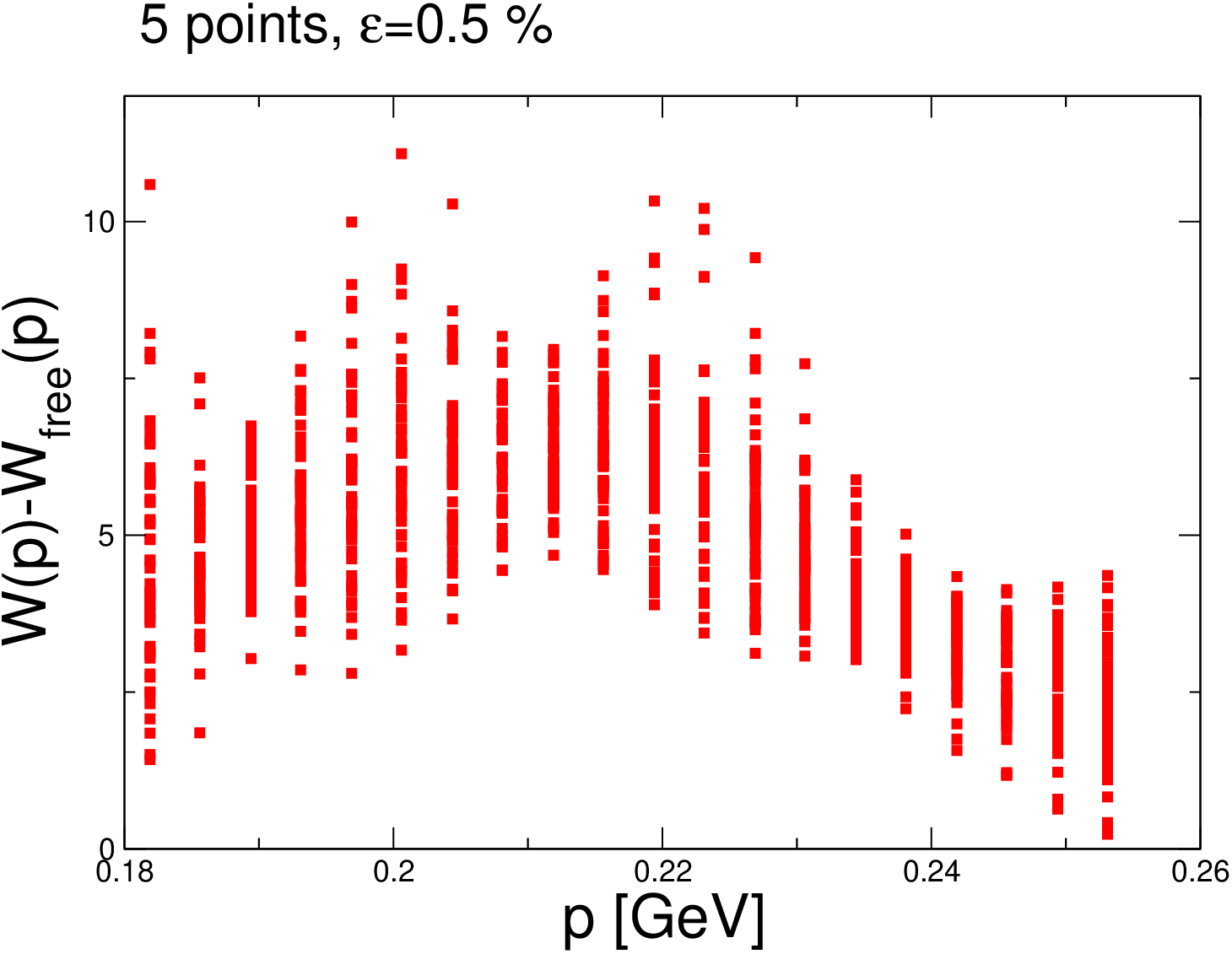}
\vspace*{.9cm}

\includegraphics[width=5.0cm]{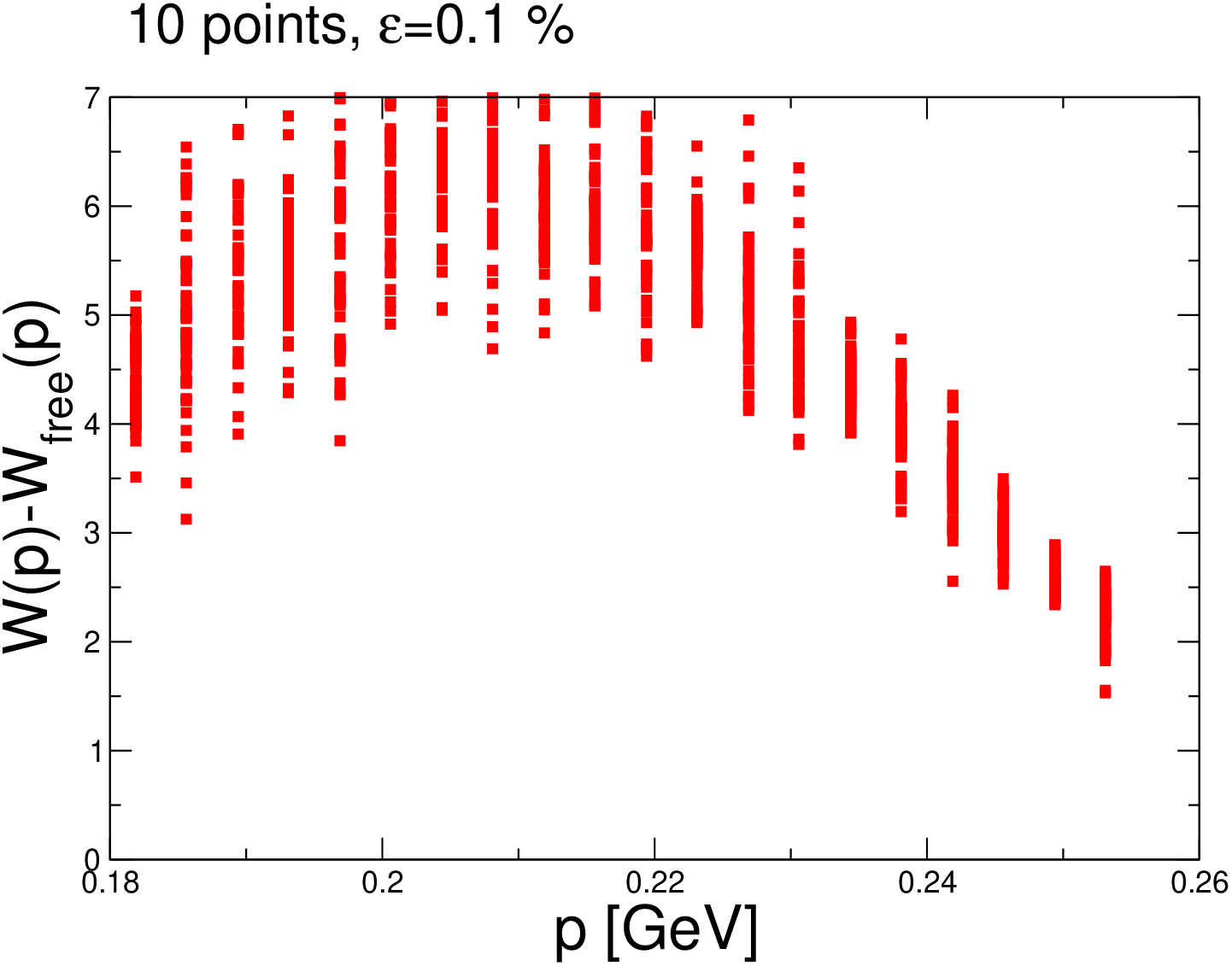}\hspace*{.2cm}
\includegraphics[width=5.0cm]{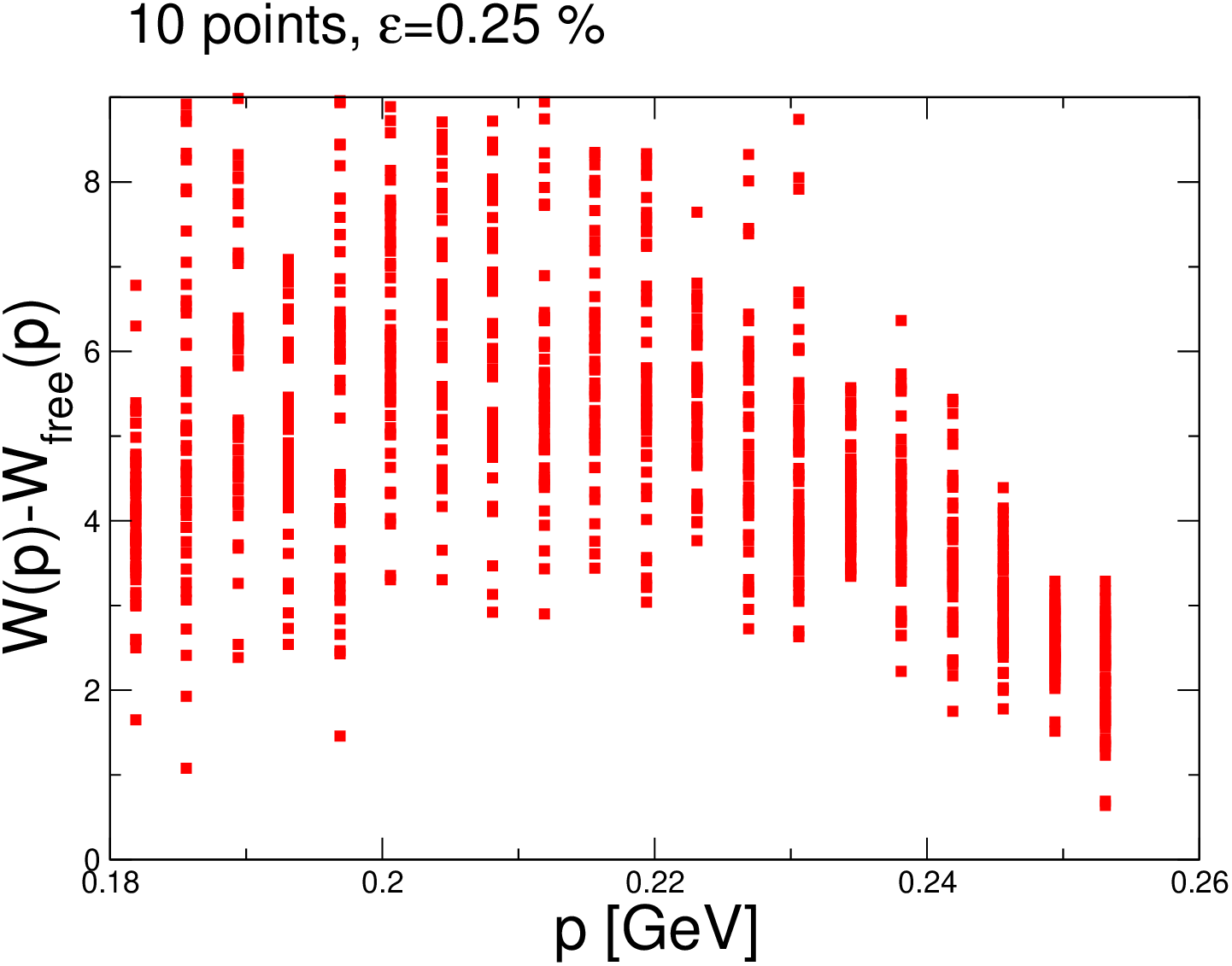}\hspace*{.2cm}
\includegraphics[width=5.0cm]{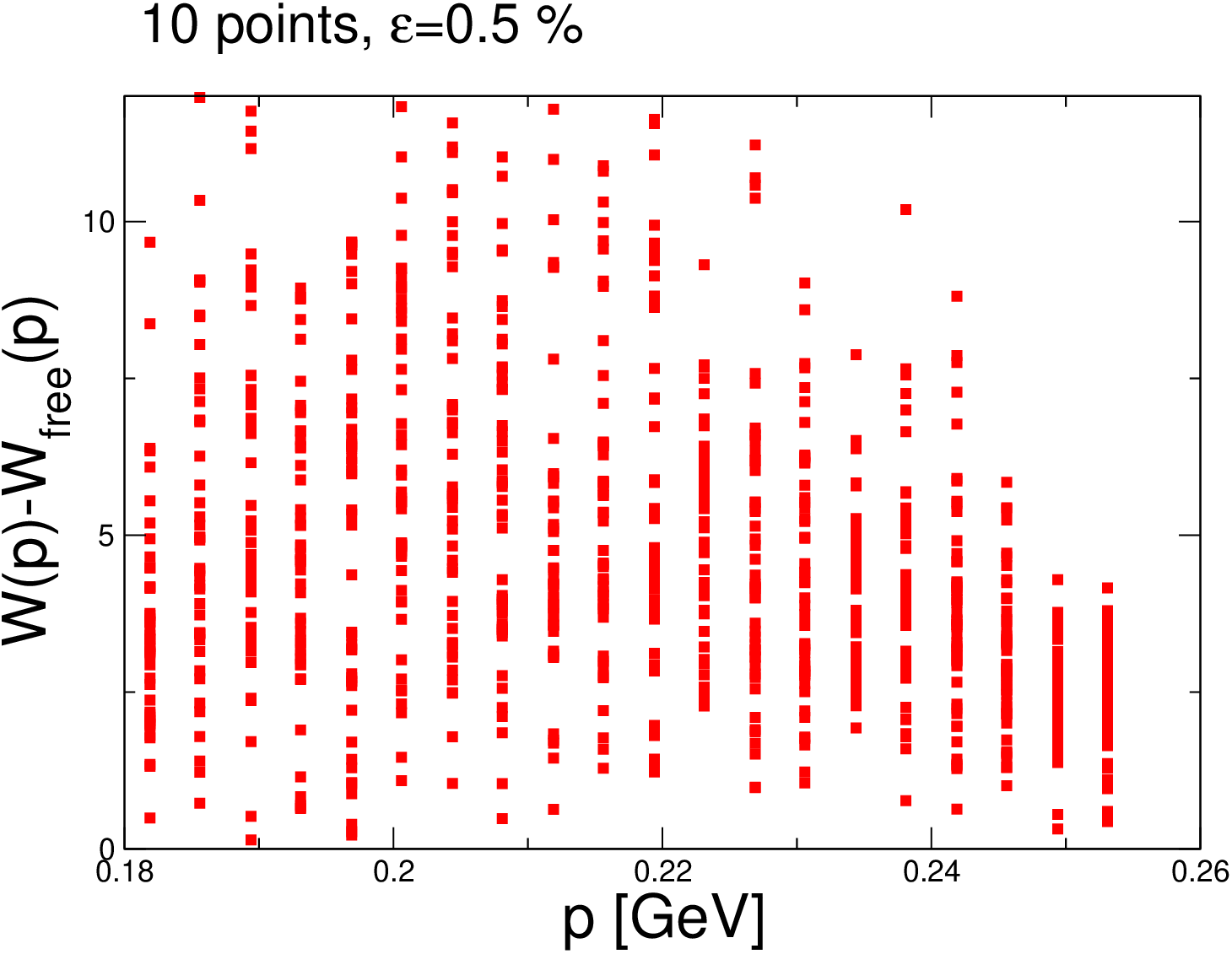}
\end{center}
\caption{Probability distributions, obtained from the data that contain
errors (see the text for the details). The central data points are the same
as in Fig.~\ref{fig:improving}. As seen from the figure, 
the resonance structure is
effectively washed out already at a relative error of $0.5 \%$ in the data
or even earlier. }
\label{fig:errors}
\end{figure}

\begin{figure}[ht]
\begin{center}
\includegraphics[width=6.0cm]{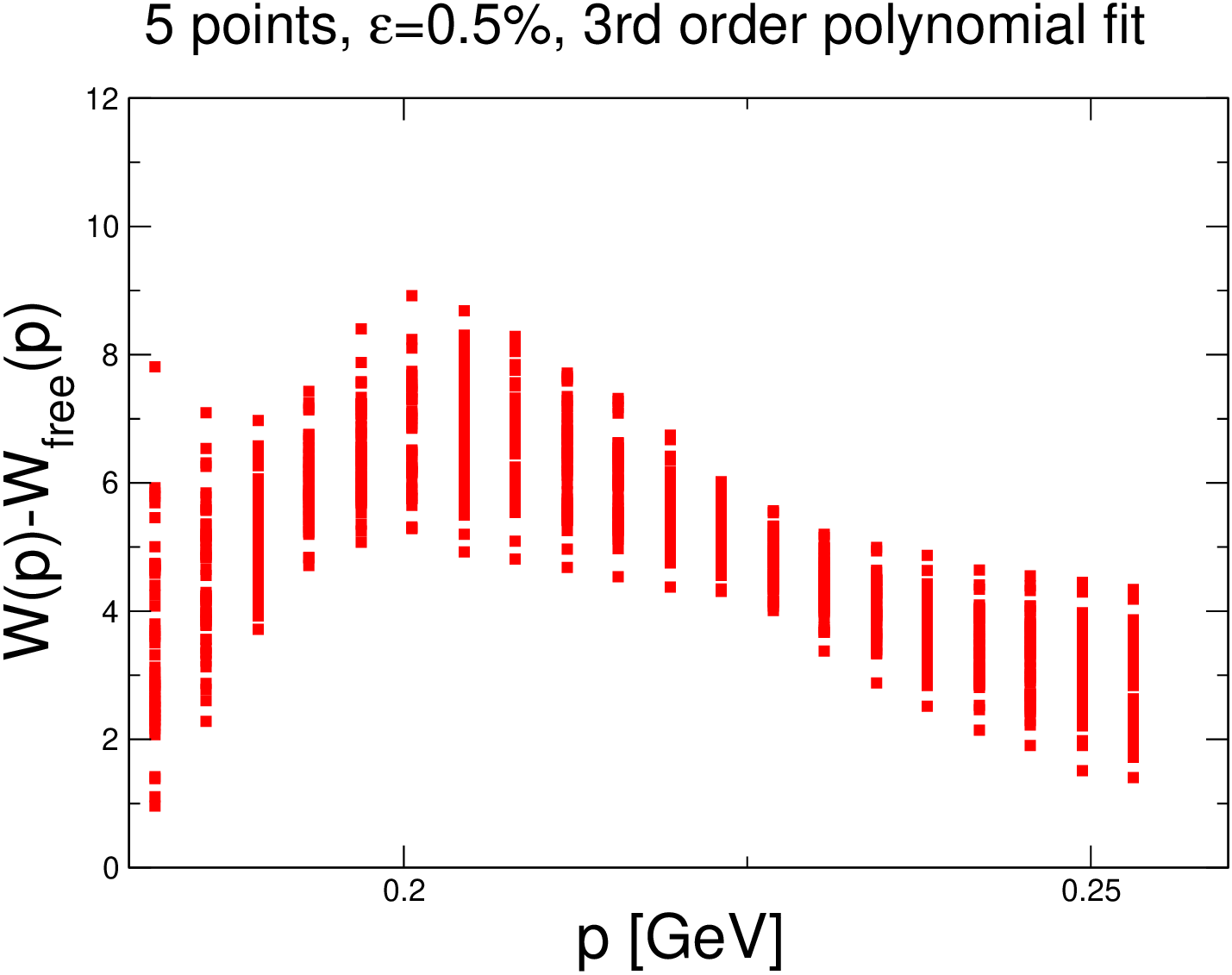}\hspace*{.2cm}
\includegraphics[width=6.0cm]{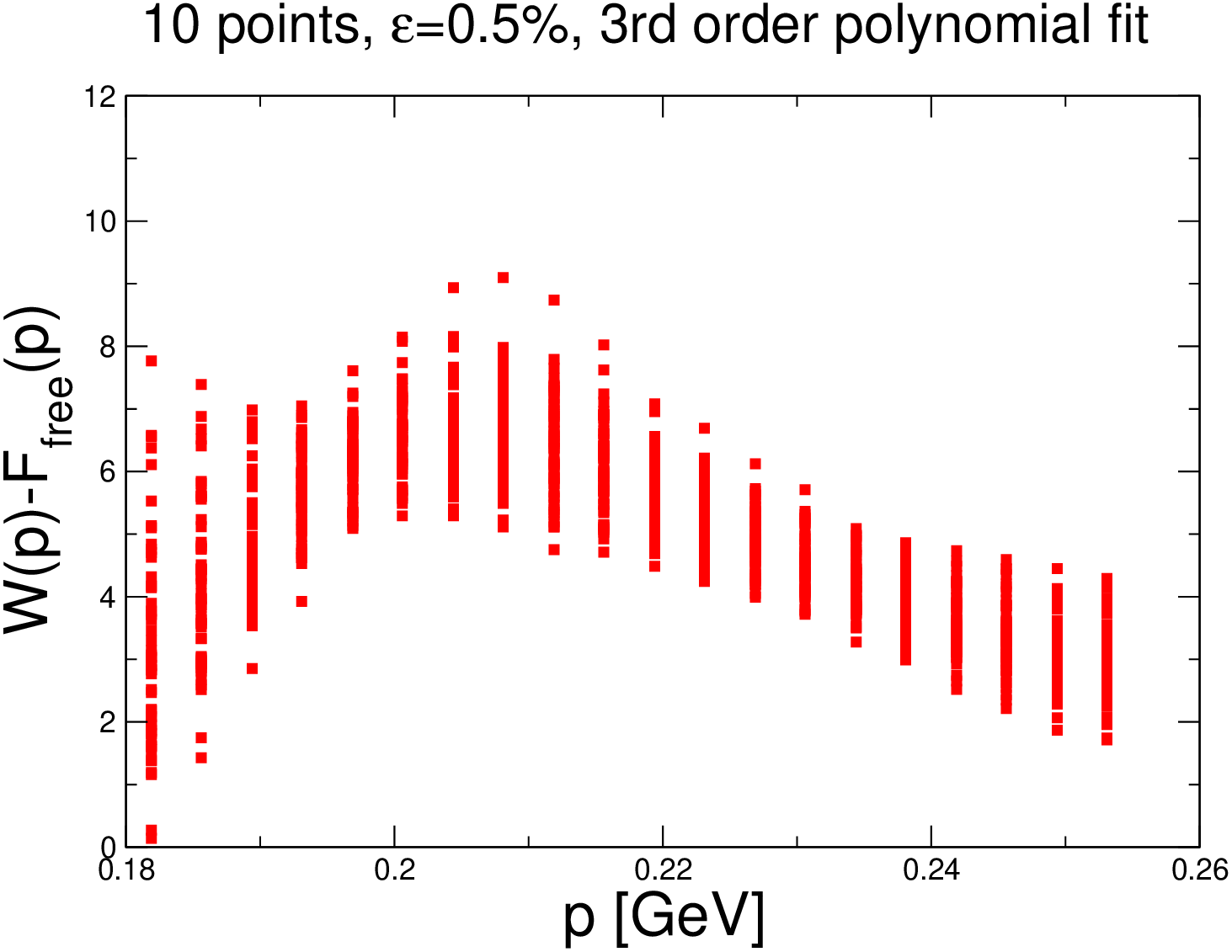}
\end{center}
\caption{The same as in Fig.~\ref{fig:errors} at a relative error 
$\varepsilon=0.5\%$. A polynomial fit to the random data was used
prior to producing the probability distribution. The improvement as compared
to Fig.~\ref{fig:errors} is clearly visible. The peak is washed out approximately at $\varepsilon\simeq 0.75\%\cdots 1\%$.}

\label{fig:polynom}
\end{figure}

We start from the data equidistantly distributed in the interval 
$M_\pi L\in [1.9,4.5]$ (5 or 10 data points, as in Fig.~\ref{fig:errors}).
From each data point $E_i=E(L_i)$ we further produce a statistical 
sample of 50 data points at a same $L_i$, which are normally distributed
around the central value $E_i$ with the standard deviation 
$\sigma_i=\varepsilon E_i$. The figure~\ref{fig:errors} shows the probability
distributions obtained from these randomly produced data, for three
different values of $\varepsilon=0.1\%,0.25\%,0.5\%$. Spline 
interpolation is used between data points. It is evident that
 the nice resonance
structure, which was seen in the probability distribution (data with no errors),
is washed out already at quite small values of the relative error assigned. 
Interestingly enough, the increase of the number of data points does not lead
to an improved accuracy. The reason for this is that
we treat the neighboring data points to be statistically independent.
If the distance between the neighboring points is decreased, the
fluctuations in the derivative of the eigenvalues increase
and this leads to the increase of the statistical noise in the probability
distribution (see Fig.~\ref{fig:errors}).

These statistical fluctuations can be suppressed to some extent if 
we perform a smooth  (e.g. polynomial) interpolation
of random data prior to calculation of the probability distribution.
This is demonstrated in Fig.~\ref{fig:polynom} where the random data 
on the spectrum were first fitted as
$p(L)=a_0+a_1L+a_2L^2+a_3L^3$ prior to producing the probability distribution.
The situation clearly improves, especially in the case of 10 data points.
Still, from Figs.~\ref{fig:errors} and \ref{fig:polynom}
one has to conclude
that a {\em very accurate} measurement of the spectrum is 
indeed needed
to reliably extract the properties of the $\Delta$-resonance.

\section{Conclusions}
\label{sec:concl}

\begin{itemize}

\item[i)]
Within the covariant non-relativistic effective field theory~\cite{cuspwe}
we have derived L\"uscher's formula for scattering of  spin-1/2
and spin-$0$ particles. The partial-wave expansion is performed, and the 
cubic symmetry on the lattice is used to reduce the resulting matrix
equations. 

\item[ii)]
The notion of probability distribution for the finite-volume
spectrum of the Hamiltonian is introduced. It is shown that near the
resonance energy the probability distribution behaves similar to the
scattering cross section in the infinite volume: it produces a Breit-Wigner
peak at the resonance energy with the same width.

\item[iii)]
The probability distribution, which is directly constructed from the energy 
levels, does not contain any prior bias. For this reason, the analysis carried
out with the use of this method can be used to judge whether a clean
extraction of the resonance parameters from the available data is possible.
 
\item[iv)]
The probability distribution does not carry more or less physical information
than L\"uscher's formula. 
The advantage of the probability distribution is its visual transparency.
The choice of the method in the actual analysis
is dictated by  convenience.

\item[v)]
In the present paper we apply the method of probability distribution
to the case of the $\Delta$-resonance.
We observe that the distribution {\em after subtracting the background 
corresponding to the free motion of the $\pi N$ pair} develops a nice
resonance structure in accordance with the exact prediction based on 
L\"uscher's formula -- even though the avoided level crossing is completely 
washed out.

\item[vi)]
It is possible to achieve a satisfactory description of the resonance
position and shape even with few data points, provided they are chosen
close enough to the resonance and are measured very accurately.
Measurement of only the ground state suffices (inclusion of the
excited levels provides additional check on the results). To conclude,
the results of the paper clearly demonstrate that the extraction of the 
resonance parameters from the measurement of the finite-volume energy spectrum
by using L\"uscher's method is indeed a feasible although difficult task.  

\end{itemize}

\vfill\eject

\noindent{\large {\bf Acknowledgments}}

\smallskip\noindent
We are grateful to J.~Gasser, H.~Krebs, J.~Negele,  F.~Niedermayer,
H.~Petry, G.~Schierholz, U.-J.~Wiese and U.~Wenger for 
useful discussions.


\renewcommand{\thefigure}{\thesection.\arabic{figure}}
\renewcommand{\thetable}{\thesection.\arabic{table}}
\renewcommand{\theequation}{\thesection.\arabic{equation}}

\appendix

\setcounter{equation}{0}
\setcounter{figure}{0}
\setcounter{table}{0}

\section{Basis vectors of the cubic group}
\label{app:basisvectors}

In this appendix we collect the basic formulae for the representations
of the cubic group, which are needed for partial diagonalization  of L\"uscher's formula. 

Consider first shortly the case without spin.
The group $O$ of the symmetries of the 3-dimensional cube has 24 elements 
(no reflections included),
which fall into 5 conjugacy classes: $I$ (identity), $3C_2$, $8C_3$,
$6C_4$ and $6C_2'$ (see, e.g.~\cite{Johnson,Mandula}). 
We find it convenient to parametrize the group elements
by specifying three Euler angles $\alpha,\beta,\gamma$.
Alternatively, the element can be parametrized, e.g., by
specifying the axis ${\bf n}$ and the rotation angle $\omega$.
The table~\ref{tab:parameters} collects the values of the group parameters.

\begin{table}[ht]

\renewcommand{\arraystretch}{1.5}

\begin{center}
\begin{tabular}{|rcrrrrr|}\hline
 Class &$i$   &${\bf n}$  & $\omega$   & $\alpha$ & $\beta$  & $\gamma$   \\
\hline
 $I$   & 1    & any       & $0$        & $0$      & $0$      & $0$        \\ \hline
 $8C_3$& 2    &$(1,1,1)$  & $-2\pi/3$ & $-\pi /2$& $-\pi /2$& $0$        \\
       & 3    &$(1,1,1)$  & $2\pi/3$  & $0$      & $\pi /2$ & $\pi /2$   \\
       & 4    &$(-1,1,1)$ & $-2\pi/3$ & $0$      & $-\pi /2$& $-\pi /2$  \\
       & 5    &$(-1,1,1)$ & $2\pi/3$  & $\pi /2$ & $\pi /2$ & $0$        \\
       & 6    &$(-1,-1,1)$& $-2\pi/3$ & $-\pi /2$& $\pi /2$ & $0$        \\
       & 7    &$(-1,-1,1)$& $2\pi/3$  & $0$      & $-\pi /2$& $\pi /2$   \\
       & 8    &$(1,-1,1)$ & $-2\pi/3$ & $0$      & $\pi /2$ & $-\pi /2$  \\
       & 9    &$(1,-1,1)$ & $2\pi/3$  & $\pi /2$ & $-\pi /2$& $0$        \\ \hline
 $6C_4$& 10   &$(1,0,0)$  & $-\pi /2$  & $-\pi /2$& $-\pi /2$& $\pi /2$   \\
       & 11   &$(1,0,0)$  & $\pi /2$   & $\pi /2$ & $-\pi /2$& $-\pi /2$  \\
       & 12   &$(0,1,0)$  & $-\pi /2$  & $0$      & $-\pi /2$& $0$        \\
       & 13   &$(0,1,0)$  & $\pi /2$   & $0$      & $\pi /2$ & $0$        \\
       & 14   &$(0,0,1)$  &  $-\pi /2$ & $-\pi /2$& $0$      & $0$        \\
       & 15   &$(0,0,1)$  &  $\pi /2$  & $\pi /2$ & $0$      & $0$        \\ \hline
$6C'_2$& 16   &$(0,1,1)$  &  $-\pi$    & $-\pi /2$& $-\pi /2$& $-\pi /2$  \\
       & 17   &$(0,-1,1)$ &  $-\pi$    & $-\pi /2$& $\pi /2$ & $-\pi /2$  \\
       & 18   &$(1,1,0)$  &  $-\pi$    & $-\pi /2$& $-\pi$   & $0$        \\
       & 19   &$(1,-1,0)$ &  $-\pi$    & $0$      & $\pi$    & $-\pi /2$  \\
       & 20   &$(1,0,1)$  &  $-\pi$    & $0$      & $\pi /2$ & $-\pi$     \\
       & 21   &$(-1,0,1)$ &  $-\pi$    & $0$      & $-\pi /2$& $-\pi$     \\ \hline
 $3C_2$& 22   &$(1,0,0)$  &  $-\pi$    & $\pi$    & $\pi$    & $0$        \\
       & 23   &$(0,1,0)$  &  $-\pi$    & $0$    & $-\pi$   & $0$        \\
       & 24   &$(0,0,1)$  &  $-\pi$    & $0$    & $0$      & $-\pi$     \\
\hline
\end{tabular}
\end{center}
\caption{Parameterization of the elements of cubic group. The vector ${\bf n}$ should be normalized to unity.}
\label{tab:parameters}
\end{table}

The irreducible representations of the rotation group $D^L$, $L=0,1,\cdots$ 
are defined in the $2L+1$-dimensional space spanned on the basis 
vectors $|LM\rangle$. These representations are reducible
under the cubic group $O$ and can be decomposed into the irreducible 
representations of the latter denoted by $\Gamma=A_1,A_2,E,T_1~\mbox{and}~T_2$.
The dimension of these representations $N(\Gamma)$ is equal 
to 1,1,2,3,3, respectively.

In order to construct the basis of the irreducible representations, we
consider the linear operator $P_{\alpha\beta}^{\Gamma,L}$, whose matrix
elements in the space spanned by the vectors $|LM\rangle$ are given by
\eq\label{eq:proj}
(P_{\alpha\beta}^{\Gamma,L})_{MM'}=\sum_{i=1}^{24}
(R_i^\Gamma)^*_{\alpha\beta}\,D^L_{MM'}(\alpha_i,\beta_i,\gamma_i)\, ,
\en
where $D^L_{MM'}(\alpha_i,\beta_i,\gamma_i)$ are Wigner $D$-functions, and
$(R_i^\Gamma)_{\alpha\beta}$, $\alpha,\beta=1\cdots N(\Gamma)$
denote the matrices of the irreducible representations of cubic group

\begin{itemize}

\item[$A_1$:] $A_1$ is the trivial 1-dimensional representation $R_i=1$.

\item[$A_2$:] $R_i=-1$ for the conjugacy classes $6C_4$ and $6C_2'$, $R_i=1$
otherwise.

\item[$E$:] The matrices in this representation are two-dimensional and real:
\eq
R_i&=&{\bf 1}\quad\mbox{for}\quad i=1,22,23,24,
\nonumber\\[2mm]
R_i&=&\sigma_3\quad\mbox{for}\quad i=14,15,18,19,
\nonumber\\[2mm]
R_i&=&-\cos\frac{\pi}{3}\,{\bf 1}+i\sin\frac{\pi}{3}\,\sigma_2
\quad\mbox{for}\quad i=2,5,6,9,
\nonumber\\[2mm]
R_i&=&-\cos\frac{\pi}{3}\,{\bf 1}-i\sin\frac{\pi}{3}\,\sigma_2
\quad\mbox{for}\quad i=3,4,7,8,
\nonumber\\[2mm]
R_i&=&-\cos\frac{\pi}{3}\,\sigma_3-\sin\frac{\pi}{3}\,\sigma_1
\quad\mbox{for}\quad i=10,11,16,17,
\nonumber\\[2mm]
R_i&=&-\cos\frac{\pi}{3}\,\sigma_3+\sin\frac{\pi}{3}\,\sigma_1
\quad\mbox{for}\quad i=12,13,20,21.
\en
\item[$T_1$:] $(R_i)_{\alpha\beta}=\exp\biggl(-i{\bf n}^{(i)}{\bf J}\,\omega_i\biggr)_{\alpha\beta}=\cos\omega_i\delta_{\alpha\beta}
+(1-\cos\omega_i)n_\alpha^{(i)} n_\beta^{(i)} 
-\sin\omega_i\varepsilon_{\alpha\beta\gamma}n_\gamma^{(i)}$,
where $(J_{\gamma})_{\alpha\beta}=-i\varepsilon_{\alpha\beta\gamma}$
denote the group generators.

\item[$T_2$:] The matrices are the same as in the irreducible representation
$T_1$, except the change of sign for the conjugacy classes $6C_4$ and $6C_2'$.

\end{itemize}

The basis vectors of the irreducible representations are obtained by
 acting with the linear operator given by Eq.~(\ref{eq:proj}) 
{\em at a fixed $\beta$ and varying $\alpha$} on an arbitrary vector
$\phi_M$ from the space spanned by the vectors $|LM\rangle$
\eq\label{eq:basisvectors}
(e_\alpha^{\Gamma,L,\beta})_M={\cal N}\,\sum_{M'=-L}^L
(P_{\alpha\beta}^{\Gamma,L})_{MM'}\phi_{M'}\, ,\quad\quad 
\alpha=1\cdots N_\Gamma\, ,\quad\quad \beta~\mbox{fixed}\, ,
\en
where ${\cal N}$ denotes the normalization constant.
It is fixed so that the basis vectors obey the orthonormality
condition
\eq\label{eq:ONB}
\sum_M(e_{\alpha'}^{\Gamma',L,\beta})^*_M\,(e_{\alpha}^{\Gamma,L,\beta})_M
=\delta_{\alpha'\alpha}\delta_{\Gamma'\Gamma}\, .
\en
If the representation $\Gamma$ is not contained in $D^L$, the action of the
projection operator on $\phi_M$ gives 0. The equations~(\ref{eq:basisvectors})
and (\ref{eq:ONB})
do not fix a common phase of the basis vectors, belonging to the same
representation labeled with $L,\Gamma$ -- this can be freely chosen.
 If a representation $\Gamma$ enters more 
than once in $D^L$, 
an additional orthogonalization of the basis vectors, belonging
to the same representation, is necessary.
Inclusion of parity is trivial. The basis vectors are simultaneously the 
eigenvectors of $S_2$ with the eigenvalue $P=(-)^L$.

In table~\ref{tab:integer} we list the basis vectors of the irreducible
representations of the cubic group up to $L=4$, obtained from 
Eq.~(\ref{eq:basisvectors}). The phases are chosen so that after the 
partial diagonalization 
of L\"uscher's equation, the entries of table~E.2 in Ref.~\cite{Luescher_torus}
are reproduced.

Note also that our basis differs from the one given in Ref.~\cite{Altmann}
and can not be reduced to it with a single unitary transformation for all
$L$. We have also checked that the use of the basis from Ref.~\cite{Altmann}
does not lead to Eq.~(\ref{eq:Schur}): e.g., the different irreducible 
representations turn out not to be orthogonal.

Our basis vectors can be transformed into those listed in Ref.~\cite{Lee:2008fa} by unitary transformations. Since in that article the basis vectors of the same irreducible representation, belonging to different values of $L$, are not fully displayed, we can not carry out this comparison to the end.

\renewcommand{\baselinestretch}{1.4}

\begin{table}[ht]

\renewcommand{\arraystretch}{1}

\begin{center}
\begin{tabular}{|cccc|}

\hline

$\Gamma$ &$L$ & $\alpha$ & Basis vectors \\

\hline

 $A_1^+$ &$0$ & 1 & $|0,0\rangle$ \\

\hline

$T_1^-$ &$1$ & 1 &      $\frac{1}{\sqrt{2}}\,(|1,-1\rangle-|1,1\rangle)$ \\
        &    & 2 &     $ \frac{i}{\sqrt{2}}\,(|1,-1\rangle+|1,1\rangle)$ \\
        &    & 3 &                                         $|1,0\rangle$ \\

\hline

 $T_2^+$ &$2$ & 1 &     $-\frac{1}{\sqrt{2}}\,(|2,-1\rangle+|2,1\rangle)$ \\
         &    & 2 &         $\frac{i}{\sqrt{2}}\,(|2,-1\rangle-|2,1\rangle)$ \\
         &    & 3 &       $-\frac{1}{\sqrt{2}}\,(|2,-2\rangle-|2,2\rangle)$ \\

\hline

 $E^+$  &$2$  & 1 &  $|2,0\rangle$ \\
        &     & 2 & $\frac{1}{\sqrt{2}}\,(|2,-2\rangle+|2,2\rangle)$\\

\hline

 $T_1^-$ &$3$ & 1 & $\frac{\sqrt{5}}{4}\,(|3,-3\rangle-|3,3\rangle)
                        -\frac{\sqrt{3}}{4}\,(|3,-1\rangle-|3,1\rangle)$\\
        &     & 2 &$\frac{-i\sqrt{5}}{4}\,(|3,-3\rangle+|3,3\rangle)
                       -\frac{i\sqrt{3}}{4}\,(|3,-1\rangle+|3,1\rangle)$\\
        &     & 3 &$|3,0\rangle$\\

\hline

 $T_2^-$ &$3$ & 1 & $-\frac{\sqrt{3}}{4}\,(|3,-3\rangle-|3,3\rangle)
                       -\frac{\sqrt{5}}{4}\,(|3,-1\rangle-|3,1\rangle)$\\
        &     & 2 &$\frac{-i\sqrt{3}}{4}\,(|3,-3\rangle+|3,3\rangle)
                     +\frac{i\sqrt{5}}{4}\,(|3,-1\rangle+|3,1\rangle)$\\
        &     & 3 &$\frac{1}{\sqrt{2}}\,(|3,-2\rangle+|3,2\rangle)$\\

\hline

 $A_2^-$ &$3$ & 1 & $\frac{1}{\sqrt{2}}\,(|3,-2\rangle-|3,2\rangle)$\\

\hline
 $T_1^+$ &$4$ & 1 & $-\frac{1}{4}\,(|4,-3\rangle+|4,3\rangle)
                       -\frac{\sqrt{7}}{4}\,(|4,-1\rangle+|4,1\rangle)$\\
        &     & 2 &$\frac{i}{4}\,(|4,-3\rangle-|4,3\rangle)
                       -\frac{i\sqrt{7}}{4}\,(|4,-1\rangle-|4,1\rangle)$\\
        &     & 3 &$\frac {1}{\sqrt{2}}\,(|4,-4\rangle-|4,4\rangle)$\\

\hline
 $T_2^+$ &$4$ & 1 & $\frac{\sqrt{7}}{4}\,(|4,-3\rangle+|4,3\rangle)
                       -\frac{1}{4}\,(|4,-1\rangle+|4,1\rangle)$\\
        &     & 2 &$\frac{i\sqrt{7}}{4}\,(|4,-3\rangle-|4,3\rangle)
                     +\frac{i}{4}\,(|4,-1\rangle-|4,1\rangle)$\\
        &     & 3 &$\frac {1}{\sqrt{2}}\,(|4,-2\rangle-|4,2\rangle)$\\

\hline
 $E^+$   &$4$ & 1 & $-\frac{\sqrt{42}}{12}\,(|4,-4\rangle+|4,4\rangle)
                      +\frac{\sqrt{15}}{6}\,|4,0\rangle$\\
        &     & 2 &$-\frac {1}{\sqrt{2}}\,(|4,-2\rangle+|4,2\rangle)$\\

\hline
 $A_1^+$ &$4$ & 1 & $\frac{\sqrt{30}}{12}\,(|4,-4\rangle+|4,4\rangle)
              +\frac{\sqrt{21}}{6}\,|4,0\rangle$\\

\hline

\end{tabular}
\end{center}
\caption{The normalized basis of the irreducible representations of the cubic group:
integer values of the angular momentum.}
\label{tab:integer}
\end{table}

In order to include the particles with spin, one has to consider 
the double cover of $O$ denoted by $^2O$, which can be constructed by adding 
a negative identity for $\pm 2\pi$ rotations to the group $O$~\cite{Johnson}.
One ends up with a group of 48 elements divided into 8 conjugacy classes 
(see table~\ref{tab:fermion})
and, accordingly, with 8 irreducible representations. In addition
to the previously considered 5, one has 3 new representations denotes as
$\Gamma=G_1,G_2,H$ with the dimension $N(\Gamma)=2,2,4$, respectively. 
In case of the 
half-integer total momentum $J$, one needs to consider only these additional
even-dimensional representations. 

\begin{table}[t]

\renewcommand{\arraystretch}{0.99}

\begin{center}
\begin{tabular}{|lrrr|lrrr|}\hline
 Class   &$i$  &${\bf n}$  & $\omega$    & Class & $i$   & ${\bf n}$  & $\omega$ \\ \hline
$I$      & 1   & any       &  $0$        & $8C_3$& 28    & $(1,1,1)$  & $4\pi/3$ \\ \cline{1-4}
$6C_4$   & 2   &$(1,0,0)$  &  $\pi$      &       & 29    & $(-1,1,1)$ & $4\pi/3$ \\
         & 3   &$(0,1,0)$  &  $\pi$      &       & 30    & $(-1,-1,1)$& $4\pi/3$ \\
         & 4   &$(0,0,1)$  &  $\pi$      &       & 31    & $(1,-1,1)$ & $4\pi/3$ \\
         & 5   &$(1,0,0)$  &  $-\pi$     &       & 32    & $(1,1,1)$  & $-4\pi/3$ \\
         & 6   &$(0,1,0)$  &  $-\pi$     &       & 33    & $(-1,1,1)$ & $-4\pi/3$ \\
         & 7   &$(0,0,1)$  &  $-\pi$     &       & 34    & $(-1,-1,1)$& $-4\pi/3$ \\ \cline{1-4}
$6C'_8$  & 8   &$(1,0,0)$  &  $\pi/2$    &       & 35    & $(1,-1,1)$ & $-4\pi/3$ \\ \cline{5-8}
         & 9   &$(0,1,0)$  &  $\pi/2$    &$12C'_4$& 36    & $(0,1,1)$  & $\pi$    \\
         & 10  &$(0,0,1)$  &  $\pi/2$    &       & 37    & $(0,-1,1)$ & $\pi$    \\
         & 11  &$(1,0,0)$  &  $-\pi/2$   &       & 38    & $(1,1,0)$  & $\pi$    \\
         & 12  &$(0,1,0)$  &  $-\pi/2$   &       & 39    & $(1,-1,0)$ & $\pi$    \\
         & 13  &$(0,0,1)$  &  $-\pi/2$   &       & 40    & $(1,0,1)$  & $\pi$    \\ \cline{1-4}
$6C_8$   & 14  &$(1,0,0)$  &  $3\pi/2$   &       & 41    & $(-1,0,1)$ & $\pi$    \\
         & 15  &$(0,1,0)$  &  $3\pi/2$   &       & 42    & $(0,1,1)$  & $-\pi$   \\
         & 16  &$(0,0,1)$  &  $3\pi/2$   &       & 43    & $(0,-1,1)$ & $-\pi$   \\
         & 17  &$(1,0,0)$  &  $-3\pi/2$  &       & 44    & $(1,1,0)$  & $-\pi$   \\
         & 18  &$(0,1,0)$  &  $-3\pi/2$  &       & 45    & $(1,-1,0)$ & $-\pi$   \\
         & 19  &$(0,0,1)$  &  $-3\pi/2$  &       & 46    & $(1,0,1)$  & $-\pi$   \\ \cline{1-4}
$8C_6$   & 20  &$(1,1,1)$  &  $2\pi/3$   &       & 47    & $(-1,0,1)$  & $-\pi$   \\ \cline{5-8}
         & 21  &$(-1,1,1)$ &  $2\pi/3$   &  $J$  & 48    & any  & $2\pi$   \\ \cline{5-8}
         & 22  &$(-1,-1,1)$&  $2\pi/3$   &       &       &           &          \\
         & 23  &$(1,-1,1)$ &  $2\pi/3$   &       &       &           &          \\
         & 24  &$(1,1,1)$  &  $-2\pi/3$  &       &       &           &          \\
         & 25  &$(-1,1,1)$ &  $-2\pi/3$  &       &       &           &          \\
         & 26  &$(-1,-1,1)$&  $-2\pi/3$  &       &       &           &          \\
         & 27  &$(1,-1,1)$ &  $-2\pi/3$  &       &       &           &          \\ \hline

\end{tabular}
\end{center}
\caption{Parameterization of the elements of the double cover of the 
cubic group. The vector ${\bf n}$ should be normalized to unity.}
\label{tab:fermion}
\end{table}

The matrices of the irreducible representations $G_1,G_2,H$ are given by~\cite{Johnson}

\begin{itemize}

\item[$G_1$:] 
$(R_i)_{\alpha\beta}=\exp\biggl(-\frac{i}{2}\,{\bf n}^{(i)}
\mbox{\boldmath$\sigma$}\,\omega_i\biggr)_{\alpha\beta}=\delta_{\alpha\beta}\cos\frac{\omega_i}{2}
-i{\bf n}^{(i)}\mbox{\boldmath$\sigma$}_{\alpha\beta}\sin\frac{\omega_i}{2}$.

\item[$G_2$:] the matrices are the same except change the sign in the conjugacy
classes  $6C_8,6C_8'$ and $12C_4'$.

\item[$H$:] the matrices 
$(R_i)_{\alpha\beta}
=\exp\biggl(-i{\bf n}^{(i)}{\bf J}^{\frac{3}{2}}\,\omega_i\biggr)_{\alpha\beta}$
where ${\bf J}^{\frac{3}{2}}_{\alpha\beta}$
denote the group generators in spin-3/2 case.

\end{itemize}

The counterpart of Eq.~(\ref{eq:proj}) in case of the half-integer total
 momentum $J$ is
\eq\label{eq:proj_fermion}
(P_{\alpha\beta}^{\Gamma,J})_{MM'}=\sum_{i=1}^{48}
(R_i^\Gamma)^*_{\alpha\beta}\,D^J_{MM'}(\alpha_i,\beta_i,\gamma_i)\, .
\en
Acting with the linear operator $P_{\alpha\beta}^{\Gamma,J}$
 on an arbitrary linear combination of the basis vectors 
$|JM\rangle^\pm$
defined in section~\ref{sec:cubic}, we obtain the basis of the irreducible
representations $G_1,G_2$ and $H$. Inclusion of parity is again trivial, since
the basis vectors $|JM\rangle^\pm$ are the eigenvectors of parity, with the
eigenvalue $P=\pm 1=(-)^L$. Up to the value $J=\frac{7}{2}$, these basis
vectors are listed in table~\ref{tab:fermionbasis}

\begin{table}
\begin{center}
\begin{tabular}{|cccc|}\hline
$\Gamma^\pm$  &$J$   &$\alpha$& Basis vectors       \\\hline
$G1^{\pm}$&$1/2$ &   1    & $\vert  \frac{1}{2}~\frac{1}{2} \rangle^\pm$ \\
          &      &   2    & $\vert \frac{1}{2}~-\frac{1}{2} \rangle^\pm$       \\ \hline
$G1^{\pm}$&$7/2$ &   1    & 
$\frac{\sqrt{15}}{6}\,\vert \frac{7}{2}~-\frac{7}{2} \rangle^\pm 
+\frac{\sqrt{21}}{6}\,\vert \frac{7}{2}~\frac{1}{2} \rangle^\pm    $\\
          &      &   2    & 
$\frac{-\sqrt{21}}{6}\,\vert \frac{7}{2}~-\frac{1}{2} \rangle^\pm 
                                     
-\frac{\sqrt{15}}{6}\,\vert \frac{7}{2}~\frac{7}{2} \rangle^\pm  $\\
\hline
$G2^{\pm}$&$5/2$ &   1    & 
$\frac{\sqrt{30}}{6}\,\vert \frac{5}{2}~-\frac{3}{2} \rangle^\pm          
-\frac{\sqrt{6}}{6}\,\vert \frac{5}{2}~\frac{5}{2} \rangle^\pm        $\\
          &      &   2    & 
$-\frac{\sqrt{6}}{6}\,\vert \frac{5}{2}~-\frac{5}{2} \rangle^\pm          
+\frac{\sqrt{30}}{6}\,\vert \frac{5}{2}~\frac{3}{2} \rangle^\pm        $\\
\hline
$G2^{\pm}$&$7/2$ &   1    & 
$-\frac{1}{2}\,\vert \frac{7}{2}~-\frac{3}{2} \rangle^\pm          
+ \frac{\sqrt{3}}{2}\,\vert \frac{7}{2}~\frac{5}{2} \rangle^\pm  $\\
          &      &   2    & 
$-\frac{\sqrt{3}}{2}\, \vert \frac{7}{2}~-\frac{5}{2} \rangle^\pm          
+\frac{1}{2}\,\vert \frac{7}{2}~\frac{3}{2} \rangle^\pm $\\
\hline
$H^{\pm}$ &$3/2$ &   1    & $\vert \frac{3}{2}~\frac{3}{2} \rangle^\pm$\\
          &      &   2    & $\vert \frac{3}{2}~\frac{1}{2} \rangle^\pm$\\
          &      &   3    & $\vert \frac{3}{2}~-\frac{1}{2} \rangle^\pm$\\
          &      &   4    & $\vert \frac{3}{2}~-\frac{3}{2} \rangle^\pm$\\
\hline
 $H^{\pm}$&$5/2$ &   1    & 
$\frac{-\sqrt{30}}{6}\,\vert \frac{5}{2}~-\frac{5}{2} \rangle^\pm          
-\frac{\sqrt{6}}{6}\,\vert \frac{5}{2}~\frac{3}{2} \rangle^\pm        $\\
          &      &   2    & $\vert \frac{5}{2}~\frac{1}{2} \rangle^\pm$\\
          &      &   3    & $-~\vert \frac{5}{2}~-\frac{1}{2} \rangle^\pm$\\
          &      &   4    & 
$\frac{\sqrt{6}}{6}\,\vert \frac{5}{2}~-\frac{3}{2} \rangle^\pm          
+ \frac{\sqrt{30}}{6}\,\vert \frac{5}{2}~\frac{5}{2} \rangle^\pm      $\\
\hline
$H^{\pm}$ &$7/2$ &   1    & 
$\frac{1}{2}\,\vert \frac{7}{2}~-\frac{5}{2} \rangle^\pm          
+ \frac{\sqrt{3}}{2}\,\vert \frac{7}{2}~\frac{3}{2} \rangle^\pm        $\\
          &      &   2    & 
$\frac{\sqrt{21}}{6}\,\vert \frac{7}{2}~-\frac{7}{2} \rangle^\pm          
- \frac{\sqrt{15}}{6}\,\vert \frac{7}{2}~\frac{1}{2} \rangle^\pm        $\\
          &      &   3    & 
$-\frac{\sqrt{15}}{6}\, \vert \frac{7}{2}~-\frac{1}{2} \rangle^\pm          
+ \frac{\sqrt{21}}{6}\, \vert \frac{7}{2}~\frac{7}{2} \rangle^\pm         $\\
          &      &   4    & 
$\frac{\sqrt{3}}{2}\,\vert \frac{7}{2}~-\frac{3}{2} \rangle^\pm          
+\frac{1}{2}\,\vert \frac{7}{2}~\frac{5}{2} \rangle^\pm       $\\
\hline
\end{tabular}
\end{center}
\caption{Basis functions of irreducible representations of $^2O$ in terms of the functions $\vert  J M\rangle^\pm$ defined in section~\ref{sec:cubic}.}
\label{tab:fermionbasis}
\end{table}

Using this basis to partially diagonalize L\"uscher's equation, we finally arrive
at the results displayed in table~\ref{tab:M_nl}.

\end{document}